\documentclass[12pt]{iopart}
\usepackage{subfigure}
\usepackage[sort&compress, numbers]{natbib}
\usepackage[utf8]{inputenc}
\usepackage{subcaption}
\usepackage{caption}
\usepackage{comment}
\usepackage{bbold} 
\usepackage{braket}
\usepackage{balance}
\usepackage{physics}
\usepackage{dcolumn}
\usepackage{bm}
\usepackage[unicode=true,bookmarks=true,bookmarksnumbered=false,bookmarksopen=false,breaklinks=false,pdfborder={0 0 1},backref=false,colorlinks=true,linkcolor=cyan,citecolor=cyan]{hyperref}
\usepackage{mathdots}
\usepackage[]{graphics,graphicx,epsfig}
\usepackage{mathtools}
\usepackage{csquotes}
\MakeOuterQuote{"}
\usepackage{epstopdf}
\usepackage{adjustbox}
\usepackage{diagbox}
\usepackage{color,soul}

\def\bea{\begin{eqnarray}}
\def\eea{\end{eqnarray}}

\usepackage[x11names,table]{xcolor}

\usepackage{hyperref}

\begin{document}

\title[\fontsize{10}{12}\selectfont Towards Arbitrary QUBO Optimization: Analysis of Classical and Quantum-Activated FNNs]{Towards Arbitrary QUBO Optimization: \\ Analysis of Classical and Quantum-Activated Feedforward Neural Networks}
\author{Chia-Tso Lai$^{1,3,*}$, Carsten Blank$^{2,\dag}$, Peter Schmelcher$^{3}$ and Rick Mukherjee$^{3,\ddag}$ } 

\address{%
 $^1$Fakultät für Physik und Erdsystemwissenschaften, Universität Leipzig,
 Linnéstraße 5, 04103 Leipzig, Germany
}%

\address{%
 $^2$Data Cybernetics,
 86899, Landsberg, Germany
}%

\address{%
 $^3$Zentrum f\"ur Optische Quantentechnologien, Universit\"at Hamburg, Luruper Chaussee 149, 22761 Hamburg, Germany
}%

\ead{\mailto{$^*$chiatsolai@gmail.com}, \mailto{$^{\dag}$blank@data-cybernetics.com}, and \mailto{$^{\ddag}$rick.mukherjee@physnet.uni-hamburg.de}}

\begin{abstract}
Quadratic Unconstrained Binary Optimization (QUBO) is at the heart of many industries and academic fields such as logistics, supply chain, finance, pharmaceutical science, chemistry, IT, and energy sectors, among others \cite{bangert2012optimization}. These problems typically involve optimizing a large number of binary variables, which makes finding exact solutions exponentially more difficult. Consequently, most QUBO problems are classified as NP-hard \cite{garey1979computers,lucas2014ising}. To address this challenge, we developed a powerful feedforward neural network (FNN) optimizer for arbitrary QUBO problems. In this work, we demonstrate that the FNN optimizer can provide high-quality approximate solutions for large problems, including dense 80-variable weighted MaxCut and random QUBOs, achieving an average accuracy of over 99\% in less than 1.1 seconds on an 8-core CPU. Additionally, the FNN optimizer outperformed the Gurobi optimizer \cite{gurobi} by 72\% on 200-variable random QUBO problems within a 100-second computation time limit, exhibiting strong potential for real-time optimization tasks. Building on this model, we explored the novel approach of integrating FNNs with a quantum annealer-based activation function to create a quantum-classical encoder-decoder (QCED) optimizer, aiming to further enhance the performance of FNNs in QUBO optimization.  
\end{abstract}
%
\vspace{2pc}
\maketitle

\section{Introduction}
\label{introduction}

Quadratic unconstrained binary optimization (QUBO) is an intensively studied class of optimization problems \cite{glover2019quantum,lewis2017quadratic}. The significance of QUBO can be attested by its appearance in different real-world optimization problems such as MaxCut \cite{commander2009maximum} and maximum weighted independent set (MWIS) \cite{basagni2001finding} in resource allocation, traveling salesman problem (TSP) \cite{matai2010traveling} in logistics, portfolio optimization \cite{black1992global}, and workflow scheduling \cite{yu2008workflow}. Many of these QUBO problems are considered NP-hard and a myriad of classical and quantum algorithms were developed for them \cite{nambu2022rejection,nakano2023diverse,glover1993user,kirkpatrick1983optimization,guerreschi2021solving,morrison2016branch,nabli2009overview,schuetz2022combinatorial,gabor2020insights,gurobi,merz1999genetic,boros2007local,ohzeki2011quantum,kadowaki1998quantum,farhi2000quantum,farhi2014quantum,grover1996fast,venegas2012quantum,ebadi2022quantum,motta2020determining,del2013shortcuts,moll2018quantum,egger2021warm,orus2019quantum,yarkoni2022quantum,abbas2023quantum,boyd2007branch,narendra1977branch}. Some classical optimizers, such as Gurobi \cite{gurobi} and CPLEX \cite{cplex2009v12}, aim to solve QUBO problems exactly. These two robust solvers are both predicated on the branch-and-bound algorithm \cite{morrison2016branch,gurobi,boyd2007branch,narendra1977branch}, which retrieves optimal solutions through systematic division of the original problem into subproblems. However, the branch-and-bound algorithm is not a polynomial-time method for QUBO and can become exhaustively time-consuming for larger instances. As an alternative to exact solutions, classical heuristics have been designed to find good approximate solutions for specific types of QUBO problems within a reasonable time. For example, Monte Carlo simulation \cite{nambu2022rejection}, genetic algorithms \cite{nakano2023diverse,merz1999genetic}, the tabu search algorithm \cite{glover1993user}, simulated annealing \cite{kirkpatrick1983optimization,van1987simulated}, divide-and-conquer algorithms \cite{guerreschi2021solving},  and simplex algorithms (i.e. Nelder-Mead) \cite{nabli2009overview,nelder1965simplex} are all well-sought heuristic approaches for QUBO. Nevertheless, these algorithms have no guarantee of consistent performance, and the quest for good approximate solutions could still demand substantial time or, in certain cases, may even be impossible to achieve.\\

Another interesting development in classical computing for solving QUBO problems is the use of machine learning and neural networks. Given the success of machine learning and neural networks in tackling regression and classification tasks \cite{alaa2024efficient,de2004machine,carleo2019machine}, the potential of exploiting them for QUBO has sparked some research interest. For instance, models such as unsupervised Graph Neural Networks (GNN) \cite{schuetz2022combinatorial} and supervised Autoencoders \cite{gabor2020insights} have been explored for problems such as MaxCut, MIS, TSP, and even arbitrary QUBO. In \cite{schuetz2022combinatorial}, the GNN demonstrated the ability to optimize 5-regular MaxCut graphs with up to $10^{4}$ nodes using unsupervised learning, reaching an accuracy slightly above 90\% of the currently best-published result. The autoencoder models in \cite{gabor2020insights} exhibited less than satisfying performance on 8-TSP samples with 65\% accuracy. Plus, the supervised learning method used well over 200 neurons per layer and over 10,000 datasets for training randomly generated 16-variable QUBO instances, yet it fell short of achieving an accuracy of 50\%. These neural networks might excel in handling QUBOs with specific properties, but they either have relatively high computation costs due to the supervised learning approach or cannot be generalized well to a broader scope beyond graph-based optimization. These results suggest that solving QUBO with neural networks still requires a lot of research. \\

\begin{figure}
    \centering

    \makeatletter
    \renewcommand{\thesubfigure}{}
    
    \subfigure[]
    {\includegraphics[scale=0.7]{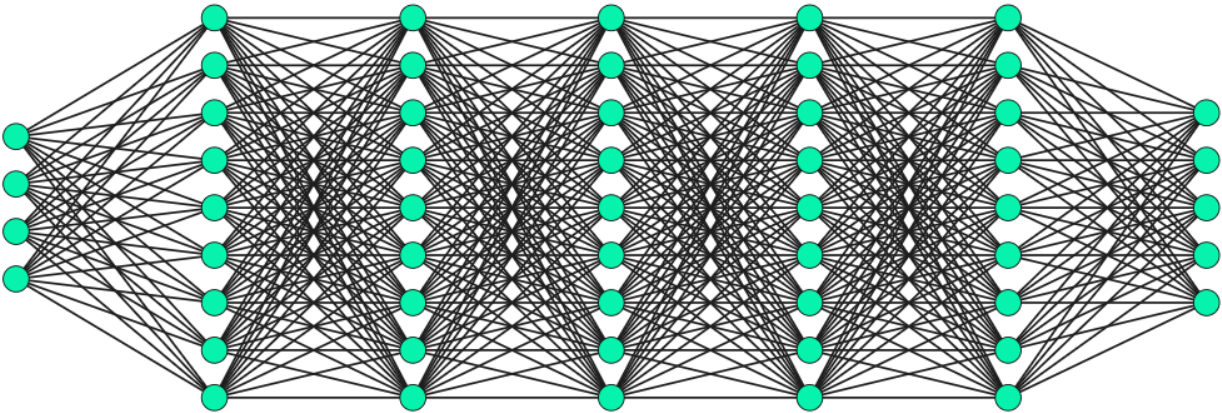}\captionsetup{font=Large}\label{fig:FNN_layout}}

    \parbox{0.4\textwidth}{\centering \large (a)}

    \subfigure[]
    {\includegraphics[width=\textwidth]{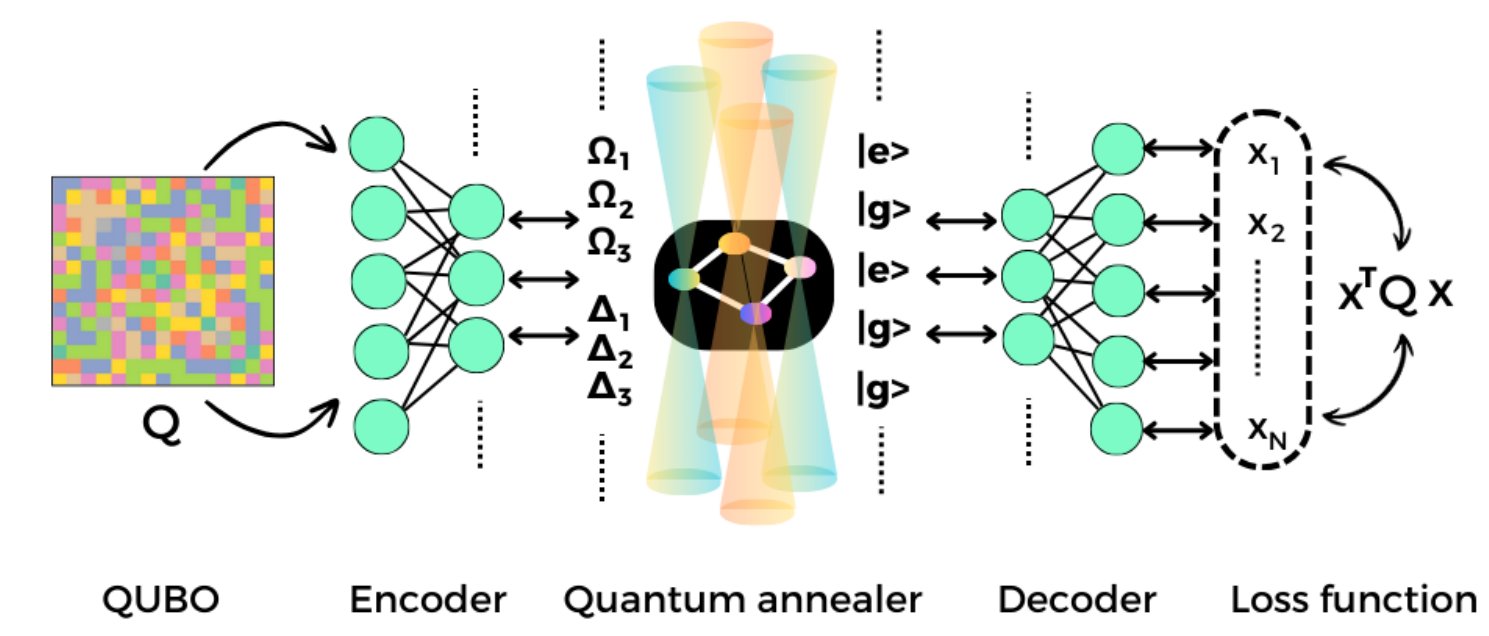}\captionsetup{font=Large}
    \label{fig:QCED_schematic}}

    \parbox{0.4\textwidth}{\centering \large (b)}

    \makeatother
    
\caption{(a) Exemplary FNN structure for solving 5-variable QUBO problems. (b) A schematic of QCED. The optimizer takes as input a QUBO matrix (Q) and consists of three major components: the encoder, the Rydberg annealer, and the decoder. In the final layer, the decoder returns a solution vector whose QUBO loss function is evaluated and minimized via gradient descent.}
\label{fig:schematic}
\end{figure}

Quantum algorithms have garnered more attention in recent years with the hope of obtaining a "quantum advantage" over the existing classical algorithms. For instance, quantum annealing \cite{ohzeki2011quantum,kadowaki1998quantum,pichler2018computational,goswami2023solving}, adiabatic quantum computing (AQC) 
 \cite{farhi2000quantum}, Grover's algorithm \cite{grover1996fast}, variational algorithms such as quantum approximate optimization algorithm (QAOA) \cite{farhi2014quantum} and variational quantum eigensolver (VQE) \cite{peruzzo2014variational}, quantum reservoir computing \cite{bravo2022quantum}, quantum walk \cite{venegas2012quantum}, quantum imaginary time evolution (QITE) \cite{motta2020determining}, and a variety of hybrid quantum-classical algorithms \cite{bosch2023neural,yang2023analog,he2024quantum}. For algorithms running on digital quantum computers, the scalability and noise of the hardware severely hinder their practicality. Conventional quantum annealing, on the other hand, faces challenges like long annealing times and variable encoding due to limitations in adiabaticity and arbitrary qubit connectivity. Given this, there are numerous potential alternatives for utilizing quantum annealers, such as leveraging the strengths of both quantum and classical computing paradigms by designing hybrid algorithms that incorporate quantum annealers \cite{bosch2023neural,yang2023analog,adachi2015application,higham2023quantum}. Very recently, alternative algorithms that are hardware-resource efficient have been explored, such as solving integer programming on a single qudit \cite{goswami2024quantum} and the TSP on a single qubit \cite{goswami2024solving}.\\

In this work, we aim to address two core questions:  First, can we exploit new neural network architectures to enhance the optimization of arbitrary QUBO problems compared to existing models? Second, is there potential to improve the neural network solver by integrating it with a quantum annealer?  Motivated by these questions, we propose using a feedforward neural network (FNN) depicted in Fig.\ref{fig:FNN_layout} to solve QUBO problems such as weighted MaxCut, MWIS, TSP, and randomly generated QUBOs. Surprisingly, despite its modest complexity with 10 layers and 756 neurons in total, the unsupervised FNN optimizer demonstrated impressive scalability and performance on 80-variable MaxCut and dense random QUBO problems, obtaining high-quality solutions with over 99\% accuracy in less than 1.1 seconds. This performance is comparable to, or even surpasses, that of powerful solvers like the Gurobi optimizer within a specified runtime limit. Moreover, the FNN's significantly reduced computation time and resources further highlights its distinctive advantage over other solvers. In an effort to facilitate the proficiency of the FNN optimizer, we explored the idea of integrating a Rydberg atom annealer into the FNN as an exotic quantum activation function. This fusion of classical neural networks with a quantum layer gave rise to the quantum-classical encoder-decoder (QCED) optimizer (Fig.\ref{fig:QCED_schematic}). Compared to FNN, QCED seemed to exhibit faster convergence and improved network initialization capabilities. However, a subsequent analysis of the parameter updates revealed that the quantum layer was effectively deactivated during training, rendering the inclusion of the annealer unjustified. Despite the quantum annealer's inactivity, many intriguing questions were raised regarding the effectiveness of quantum-classical neural networks, which pointed out a compelling direction for future research. \\

\section{Theory}
\label{theory}

In this section, we introduce the optimal FNN architecture and describe the hyperparameterization of the FNN as a QUBO optimizer. Additionally, we will cover the details of the Rydberg annealer and QCED.

\subsection{Feedforward Neural Network (FNN)}

The simplest neural network is a feedforward neural network (FNN) \cite{bebis1994feed,eldan2016power,fine2006feedforward} where multiple layers of neurons (or perceptrons) are connected by a sequence of trainable linear mappings and non-linear activation functions (i.e. ReLu, Sigmoid, tanh). Each neuron receives the inputs from the previous layer, conducts linear and non-linear transformations, and generates an output value for the next layer. Mathematically, the transformations taking place at a hidden layer (considering all neurons in that layer) can be formulated as follows:

\begin{equation}
    y = \sigma(x^{T}W+b) \ ,
\end{equation}
where $x \in \mathbb{R}^{n \times 1}$ represents the input vector from the previous layer, $W \in \mathbb{R}^{n \times m}$ is the weight matrix consisting of trainable parameters, $b \in \mathbb{R}^{1 \times m}$ denotes the bias, and $y \in \mathbb{R}^{1 \times m}$ is the output vector after applying a non-linear activation $\sigma$ to the linearly transformed input. The parameters of the network, namely $W$ and $b$, are updated using gradient descent and backpropagation algorithms given a well-defined loss function $\mathcal{L}$ at the end of the network \cite{hecht1992theory,werbos1990backpropagation,amari1993backpropagation,lecun1988theoretical}. For that, calculating the gradients is crucial in optimizing the neural network. In most cases, these gradients can be computed analytically using the chain rule for a sequence of linear and non-linear functions. As the number of neurons and layers in the model increases, the computation of gradients and hence the training becomes more and more demanding. This challenge is further compounded in supervised learning, where large volumes of input data need to be processed in batches, placing a significant burden on computational hardware. In contrast, unsupervised learning defines a data-free loss function, eliminating the need to train the model over chunks of input data, which greatly alleviates the otherwise significant workload. In the FNN and QCED optimizers, we choose unsupervised learning as the optimization paradigm by simply taking the QUBO cost function $x^{T}Qx$ as the loss without relying on any solved problem instances for the training (Fig.\ref{fig:schematic}). This allows us to efficiently optimize a larger class of QUBO problems including those for which exact solutions are unknown, as will be shown in the Results section.

\subsection{Hyperparameterization of the FNN QUBO Optimizer} 


Alongside gradient descent and backpropagation, hyperparameterization is one of the key elements contributing to the effectiveness of neural networks. Here, we detail several critical hyperparameters and features tailored for the FNN QUBO optimizer, including the pre-sampling of initial parameters, relaxation of binary variables, selection of the loss function, and the network architecture.

\paragraph{\bf Pre-sampling of initial parameters:}

A major weakness of nearly all gradient-based optimization algorithms is the curse of local minima. Since the solver navigates the loss landscape by following the principle of gradient descent, the learning process along any parameter axis comes to a halt when the corresponding gradient
vanishes. In the best possible scenario where the target function is a convex function, the solver reaches the global minimum when the gradient becomes zero. However, in most optimization problems, including QUBO, the landscape contains numerous local minima that can potentially trap the neural network during training. In light of this, more advanced gradient descent algorithms such as Adam \cite{zhang2018improved}, RMSprop \cite{zou2019sufficient} and stochastic gradient descent (SGD) \cite{zinkevich2010parallelized,ketkar2017stochastic} were proposed to mitigate the issue by either varying the learning rate based on the past gradients or incorporating randomness into the evaluation. However, such methods were designed primarily for supervised learning tasks with batches of data and are empirically not beneficial to the FNN we built for optimizing QUBO. Nevertheless, we identified a straightforward and easy-to-implement method to circumvent the premature convergence of FNN to local minima. Specifically, by adopting an unsupervised learning approach, which drastically reduces the learning time of the FNN, we can afford to try out multiple sets of initial parameters. We then train the network with these candidate starting values for a smaller number of epochs and select the best-performing one of all. Using these optimal initial parameters, FNN proceeds with the actual training for a larger number of epochs. This pre-sampling technique filters out the initial parameters whose positions in the loss landscape lead to early convergence to nearby high-energy local minima. While this approach does not guarantee finding the global optimum, it increases the chances of discovering a relatively good approximate solution.

\paragraph{\bf Relaxation of binary variables:}

When designing a neural network, it is essential to determine an appropriate output representation that aligns with the requirements of the target problem. By the definition of QUBO, binary values should be returned from the network as the solutions. However, directly mapping network outputs to discrete binary values presents a challenge, as such mappings are neither continuous nor differentiable. This discontinuity renders gradient-based optimization methods, such as gradient descent, inapplicable. As a workaround, we relax the binary constraints by allowing the variables to take continuous values between zero and one before evaluating the loss. Such relaxation can be achieved with the help of activation functions such as Sigmoid and hyperbolic tangent, 
which are commonly employed for classification tasks.\\

\paragraph{\bf Loss function:}\label{sec:loss_func}

A crucial aspect of the FNN optimizer is the choice of its loss function. Here, we examine the impacts of different loss functions on the learnability of the network. Naively, we might choose the loss as the QUBO cost function derived from the straightforward matrix multiplication $x^{T}Qx$. When expanding this product, we obtain quadratic terms for both the diagonal and off-diagonal coefficients, as indicated in the first line of Eq.(\ref{eq:qubo_loss}). The gradient of this loss function with respect to any of the variables $x_{i}$ gives a contribution from the term $2Q_{ii}x_{i}$ determined by both the diagonal coefficient and the variable $x_{i}$. However, using this form of the loss function introduces a potential issue during backpropagation because if $x_{i}$ takes a value close to zero then this part of the gradient becomes negligible, making the diagonal term irrelevant in the search for the optimal solution. To rectify this, we can instead use an equivalent QUBO cost function as the loss, which preserves the influence of all the coefficients in the derivatives. Since we are dealing with binary variables, the relation $x_{i}^{2}=x_{i}$ holds for either scenario ($x_{i}=0 \vee 1$), allowing us to linearize the quadratic terms without affecting the cost values, as shown in the second line of Eq.(\ref{eq:qubo_loss}):

\begin{equation}\label{eq:qubo_loss}
\begin{split}
x^{T}Qx &= \sum_{i}Q_{ii}x_{i}^{2}+\sum_{i}\sum_{j\neq i}Q_{ij}x_{i}x_{j} \\ &= \sum_{i}Q_{ii}x_{i}+\sum_{i}\sum_{j \neq i}Q_{ij}x_{i}x_{j} 
\end{split}
\end{equation}\\

\begin{figure}
    \centering
    \includegraphics[width=\linewidth]{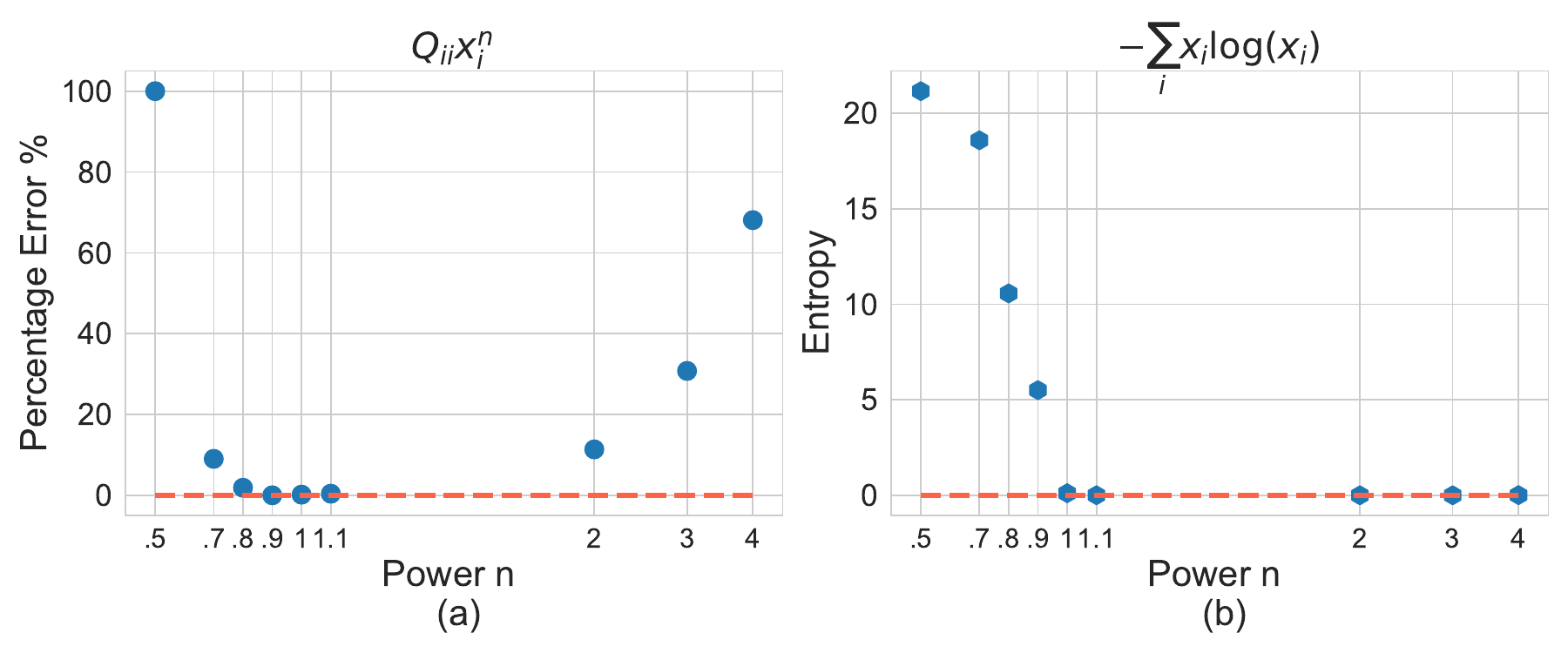}
    \caption{(a) The optimization results measured in percentage errors for different powers $n$ used in the diagonal terms. The final output values are rounded to 0 or 1 to evaluate the energy cost. (b) The Shannon entropy of the return solution vector, which quantifies how close the continuous values are to either 0 or 1.}
    \label{fig:power_loss}
\end{figure}

Empirically speaking, this adjustment proves particularly beneficial for QUBO matrices with unbalanced diagonal and off-diagonal elements, which often correspond to loss landscapes with many local minima. By introducing the linearized cost function, we reduce the likelihood of being trapped in some of these suboptimal wells. Building on the concept that $x_{i}^{2}=x_{i}$ for binary variables, we can promote this relation to more general powers $n$ such that $x_{i}^{n}=x_{i}$. To investigate how different powers of the diagonal terms impact the performance of FNN, we present the optimization results of a randomly generated QUBO problem with a known exact solution. In Fig.\ref{fig:power_loss}(a), an apparent convex pattern is observed, with the minimum error occurring somewhere between $n=0.9$ and $1$. The further the power deviates from this interval, the higher the percentage error becomes. Aside from the accuracy of the optimization, preserving the binary nature of the solution is also critical. In Fig.\ref{fig:power_loss}(b), we use Shannon entropy $-\sum_{i}x_{i}\log(x_{i})$ as a metric to assess how close the solution vector is to a true binary string. From the figure, we observe that the entropy values are larger for smaller $n$ and are almost 0 for $n \geq 1$, indicating that the relaxation of binary variables is only reasonable when $n \geq 1$.  Considering both qualities, we conclude that using $n=1$, as in the second line of Eq.(\ref{eq:qubo_loss}), provides the optimal QUBO cost function for training the FNN.

\paragraph{\bf Architecture:}

For optimization tests on MaxCut, random QUBO, and MWIS problems with 40 or fewer variables, the FNN features a 7-layer layout with 4 neurons in the input layer, $n+4$ neurons in each of the hidden layers, and $n$ neurons in the output layer to solve an $n$-variable QUBO problem. Fig.\ref{fig:FNN_layout} illustrates the FNN structure used for a 5-variable problem. For larger benchmark problems with 80 and 200 variables, a similar architecture was adopted, but the number of layers was increased to ten for improved expressivity. The input of the FNN consists of 4 trainable random numbers independent of the problem size. In FNN, all intermediate activation functions are ReLU, while the output layer employs hyperbolic tangents followed by a linear mapping to constrain the solution domain within $[0,1]^{n}$.

\subsection{Rydberg Annealer}
\label{sec:Rydberg Hamiltonian}

The cost function of a QUBO can, in theory, be mapped to the Ising model, which governs the Hamiltonian of any quantum annealer. As a result, QUBO is inherently suited for quantum annealers. The concept of quantum annealing involves encoding the cost function as the target Hamiltonian and then evolving the system from the ground state of an easy-to-prepare Hamiltonian to the ground state (or a low-energy state) of the target Hamiltonian. Since the ground state possesses the least energy, it corresponds to the optimal solution of the QUBO instance. In this work, we utilize Rydberg atom arrays as our quantum annealer of choice \cite{pichler2018computational,kim2022rydberg,lanthaler2023rydberg,qiu2020programmable,nguyen2023quantum,goswami2024loqal}.\\

Rydberg atoms can be cooled and trapped by laser beams in an experimental setup, forming an interacting many-body quantum system \cite{kaufman2021quantum,zeiher2016many,browaeys2020many,bluvstein2021controlling,ates2007many,honer2010collective}. The dynamics of a Rydberg atom array is characterized by a global Rabi frequency $\Omega$, local detunings $\Delta_{j}$, and the Van der Waals interactions $V_{jk}=C/|\bm{r}_{j}-\bm{r}_{k}|^{6}$ between pairs of atoms \cite{gaetan2009observation,johnson2010interactions,weber2017calculation,gallagher2008dipole,urban2009observation}. These laser parameters can be manipulated throughout the evolution of time to construct customized Rydberg Hamiltonians $H_{Ryd}(t)$. Specifically, the Rydberg Hamiltonian we consider incorporates a global Rabi frequency $\Omega(t)$ in the shape of a continuous step function and a set of local detunings $\Delta_{j}(t)$ tracing out linear functions in time. For training purposes, $\Omega(t)$ is parametrized by the constant value of each step $\Omega_{i} \in [0,\Omega_{max}]$ while $\Delta_{j}$ is given by the initial value $\Delta_{j}(0)$ and the slope $s_{j}$ at each atomic site $j$. Thus, the underlying Hamiltonian is given by: 
\begin{equation}\label{eq:H_ryd2}
H_{Ryd}(t) =  \frac{\hbar}{2}\Omega(t) \sum_{j}\hat{\sigma}_{j}^{x}-\hbar\sum_{j}\Delta_{j}(t)\hat{n}_{j}+\sum_{<j,k>}V_{jk}\hat{n}_{j}\hat{n}_{k} \ ,
\end{equation}
with the laser schedule $\{(\Omega(t),\Delta_{j}(t)) \mid 0 \leq t \leq T\}$:
\begin{equation}\label{eq:omega_schedule}
     \Omega(t) = \sum_{i=0}^{N-1}\frac{\bm{\omega}_{i}}{e^{-\alpha (t-t_{i})+\beta}+1} \approx \sum_{i=0}^{N-1}\Omega_{i} \chi_{(t_{i},t_{i+1}]}(t) \ ,
\end{equation}
where $\chi_{(t_{i},t_{i+1}]}(t)$ is the indicator function which returns $1$ if $t \in (t_{i},t_{i+1}]$ and 0 otherwise. For $\Omega(t)$, we approximate the N-segment discrete step function characterized by $\Omega_{i}$ and $\chi_{(t_{i},t_{i+1}]}(t)$ with a continuous function shaped by an ensemble of sufficiently steep Sigmoid functions, using the hyperparameter $\alpha \gg 1$. The coefficient $\bm{\omega}_{i}$ represents the asymptotic value of the $i^{th}$ constituent Sigmoid function and can be calculated by considering the cumulative sum of the preceding functions:
\[
     \ \ \bm{\omega}_{i}= 
\begin{cases}
    \Omega_{0},& \text{$i=0$} \\
    \Omega_{i}-\Omega_{i-1}, & \text{$i \neq 0$} 
\end{cases}
\] \\
In addition, physically-motivated boundary conditions are imposed on the Rabi frequency such that $\Omega_{0}=\Omega_{N-1}=0$. For $\Delta_{j}(t)$, the linear schedule is expressed as: 
\begin{equation}\label{eq:delta_schedule}
    \Delta_{j}(t) = \Delta_{j}(0)+s_{j}t \quad , \quad \Delta_{j}(0),s_{j}T \in [-\frac{\Delta_{max}}{2},\frac{\Delta_{max}}{2}] \ ,
\end{equation}
where we constrain $\Delta_{j}(0)$ and $s_{j}T$ within a range that ensures the experimental limitations for detunings are not violated. In the simulation, the quantum state evolves under the Hamiltonian $H_{Ryd}(t)$, after which the expectation values $\Psi_{z}^{j}(\bm{\theta})$, parametrized by a set of laser parameters $\bm{\theta}=(\Omega_{i},\Delta_{k}(0),s_{k})$, are measured in the computational basis of each atom $j$. Assuming a Rydberg atom array is initialized to the many-body state $\phi(0)$, $\Psi_{z}^{j}(\bm{\theta})$ at the time of evaluation $t=T$ reads:

\begin{equation}\label{eq:expectation}
\begin{split}
    \Psi_{z}^{j}(\bm{\theta}) &= \expval{\hat{\sigma}_{j}^{z}}{\phi(T)}\\ &=\expval{U_{Ryd}^{\dagger}(\bm{\theta})\hat{\sigma}_{j}^{z}U_{Ryd}(\bm{\theta})}{\phi(0)} \ ,
\end{split}
\end{equation}
where $U_{Ryd}(\bm{\theta}) = \exp[-\frac{i}{\hbar}\int_{0}^{T}H_{Ryd}(t)dt]$ is the time evolution operator. In the next section, we will find that these expectation values serve as outputs from the Rydberg annealer when integrated into QCED.

\subsection{Architecture of QCED}
\label{QCED}

The design of the quantum-classical encoder-decoder (QCED) optimizer consists of the following components: a QUBO matrix as the input, an encoder extracting essential information from the input via a neural network, a Rydberg annealer with trainable laser parameters, a decoder that decompresses the output from the annealer to form the solution vector, and finally, a loss function that evaluates the cost of the QUBO instance based on this solution vector. As is shown figuratively in Fig.\ref{fig:QCED_schematic}, the encoder of QCED is an FNN that takes as input the upper triangular elements of the QUBO matrix and outputs the laser parameters $(\Omega_{i},\bm{\Delta}_{j}) = (\Omega_{i},\Delta_{j}(0),s_{j})$ for the subsequent quantum layer. The encoder reduces the dimensionality of the classical data by channeling it to the quantum latent space in the form of variational quantum parameters. In the quantum layer, a parametrized Hamiltonian as described in Eqs.(\ref{eq:H_ryd2} -\ref{eq:delta_schedule}), governs the system’s time evolution, leading to expectation values measured in each atom's computational basis. These expectation values then serve as inputs for the decoding layer.  The quantum layer, fully described by the Rydberg dynamics in Eq.(\ref{eq:expectation}), can be viewed as a parametrized activation function conducting linear transformation in a high-dimensional Hilbert space followed by a projection into a lower-dimensional subspace through measurement. This suggests that the quantum layer, implemented by an analog quantum system, is likely to offer higher expressivity than a classical neural network with the same number of neurons. Following the quantum activation, the measured expectation values of the output quantum state are passed through the decoder and transformed into the desired solution vector, which is then used to calculate the loss function. The decoder, again constructed with an FNN, deciphers the information from the quantum latent space and casts it back into the solution space. To preserve the binary nature of the solution vector, the final layer is activated by hyperbolic tangent functions, which map the output values to the domain $[-1,1]$, followed by a simple rescaling to return the solution $x \in [0,1]^{n}$.\\

Similar to FNN, QCED is trained using gradient descent and backpropagation algorithms. The gradients of the quantum layer were evaluated with the finite-difference method, whose credibility has been validated by comparison with the approximated parameter-shift rule \cite{schuld2019evaluating,crooks2019gradients,wierichs2022general,banchi2021measuring}. For the parameter updates, we applied the standard gradient descent, as it demonstrated the best empirical performance. More derivations on the backpropagation of hybrid networks and the parameter-shift rule for the Rydberg annealer can be found in \ref{app:gradient_backpropogation}. The time complexity of the training process mainly depends on the number of laser parameters in the quantum layer. For a Rydberg array with $q$ atoms, we chose $q$ intermediate time steps for the waveform $\Omega(t)$ (excluding $\Omega_{0}=\Omega_{N-1}=0$). Along with two parameters $(\Delta_{j}(0),s_{j})$ for each local detuning, this results in a total of $3q$ trainable parameters. As a result, $6q$ simulations are required to compute all the gradients for one iteration of training using the finite-difference method.\\

For all optimization tests, we used only 4 qubits in QCED arranged in a square formation, leading to 12 laser parameters in the quantum layer. This minimalist configuration not only ensures that the quantum simulation is fast to execute but also enhances the likelihood of a smooth transition from noiseless simulation to actual quantum hardware. Regarding the neural network structure of QCED, the encoder and decoder are organized as follows for an $n$-variable problem: The encoder consists of 3 layers with neuron counts of [$\frac{n(n+1)}{2}+12$, $\frac{n(n+1)}{2}+12$, 12], while the decoder has 3 layers with [$n+4$, $n+4$, $n$] neurons, respectively. This results in a total of 6 classical layers plus 1 quantum layer, excluding the fixed input layer. QCED utilizes ReLU as the activation function for all hidden layers, except in the final layer of the encoder, where Sigmoid functions are used to map the laser parameters to the allowed frequency range.

\section{Results}
\label{results}

In this section, we will assess the performance of the FNN and QCED optimizers over 500 iterations of training, using 100 samples of QUBO instances that vary in both the number of variables and problem classes. To ensure a thorough and systematic evaluation, we first compare the performance and scalability of FNN and QCED by analyzing their training curves and recording their optimization results. Next, we will benchmark FNN, QCED, and other state-of-the-art classical and quantum optimizers to determine their standing in the competitive landscape of QUBO optimization. We have selected weighted MaxCut, random QUBO, MWIS, and TSP as the testing problem classes, as they encompass the realm of heavily studied graph-based optimization as well as general cases that are challenging to encode on quantum annealers.\\

Regarding the QUBO dataset, it is important to highlight that we did not restrict the degrees of the MaxCut and MWIS instances. As a result, the underlying graphs for some problems are embedded in dimensions higher than those perceivable in physical space. Consequently, the complexity of these target problems is theoretically higher compared to many k-regular graphs commonly studied in the literature.  As for random QUBO problems, we consider dense QUBO matrices with random entries in the spirit of generality. The dense, arbitrary interactions of QUBO matrices pose significant encoding challenges for quantum annealers, rendering them highly intractable for most QUBO solvers as well. More details on the data generation will be covered in \ref{app:QUBO Problems}.\\

\subsection{Optimization performance: FNN vs. QCED}

\begin{figure}
    \centering
    \includegraphics[width=\linewidth]{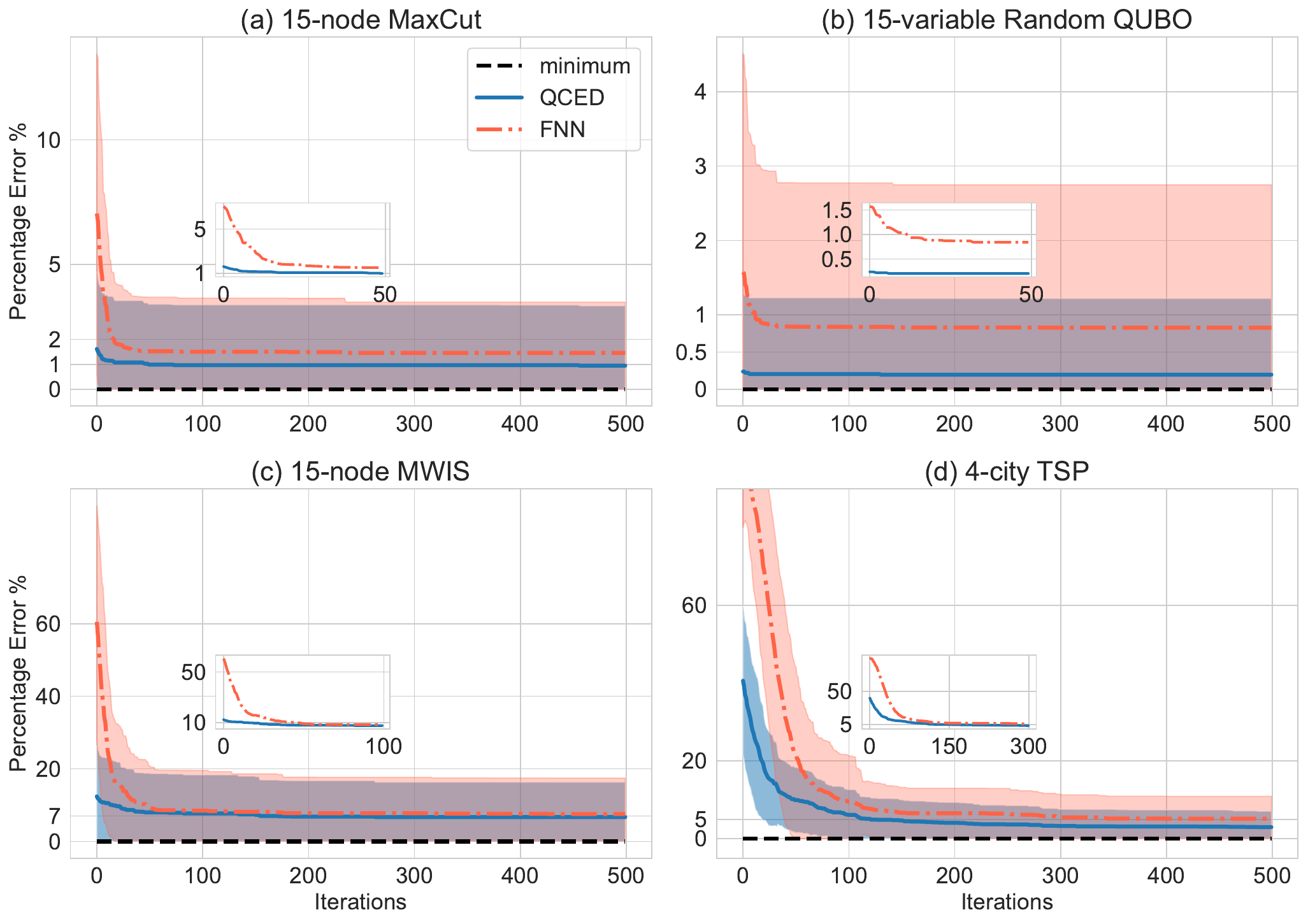}
    \caption{Learning curves of FNN and QCED for 15-node MaxCut, 15-variable random QUBO, 15-node MWIS, and 4-city TSP (16 variables). The plots display the mean values for QCED and FNN with solid and dash-dotted lines, respectively. The one-standard-deviation margins are indicated by shaded areas.}
    \label{fig:result_15_compare}
\end{figure}

As the first comparative test of optimization performance, we benchmarked 15-variable QUBO problems using both the FNN and QCED optimizer. The benchmark results are presented in Fig.\ref{fig:result_15_compare}, where the solid and dash-dotted lines denote the mean values of the percentage errors at each iteration for QCED and FNN, respectively, while the shaded regions represent one standard deviation. These statistics were calculated from 100 generated samples, and 20 rounds of pre-sampling were applied to all problems for both solvers to find good initial parameters. For 15-node MaxCut optimization, QCED obtains a better set of initial parameters and quickly converges to a rather low error of $0.95\%$, while FNN also converges after around 50 iterations but still lags behind QCED with an error of  $1.46\%$. As for 15-variable random QUBO, QCED also delivers a much better initial value, with the learning largely completed during the pre-sampling step. The error remains at a notably low $0.197\%$, outperforming the $0.826\%$ by FNN. When it comes to 15-node MWIS, both solvers suffer a significant drop in performance, with QCED reaching a suboptimal error of 6.7\% and FNN 7.53\%. However, QCED once again demonstrates an advantage in terms of parameter initialization. Lastly, in the case of 4-city TSP, the convergence of both solvers slows down compared to other problem classes, not hitting a plateau until around 300 iterations of training. At the end of the training, QCED obtains an error of $2.99\%$, while FNN yields a higher error of $5.17\%$.\\

Based on these results (Fig.\ref{fig:result_15_compare}), it seems that QCED consistently outperforms FNN. However, there are some caveats to consider. Firstly, although both QCED and FNN have 7 neural layers, the total number of classical neurons differs significantly, with 329 neurons in QCED compared to 114 in FNN for the 15-variable problems we examined. Additionally, increasing the number of layers in FNN empirically improves performance at a relatively low runtime cost, and can make its optimization results better than those achieved by QCED. It should also be noted that the overlap of the one-standard-deviation regions does not necessarily indicate equal quality of solutions because one might be comparing the results of two distinct instances in the dataset. Instead, the standard deviation only reflects the consistency of the solver's performance across different instances in this context. Regarding parameter initialization, even though we applied 20 rounds of pre-sampling to both models, we could increase the number of rounds for FNN, given that its training process is considerably faster than that of QCED. Doing so could also potentially improve the performance of FNN without significantly increasing the overall computational cost.\\

\begin{figure}
    \centering
    \includegraphics[width=\linewidth]{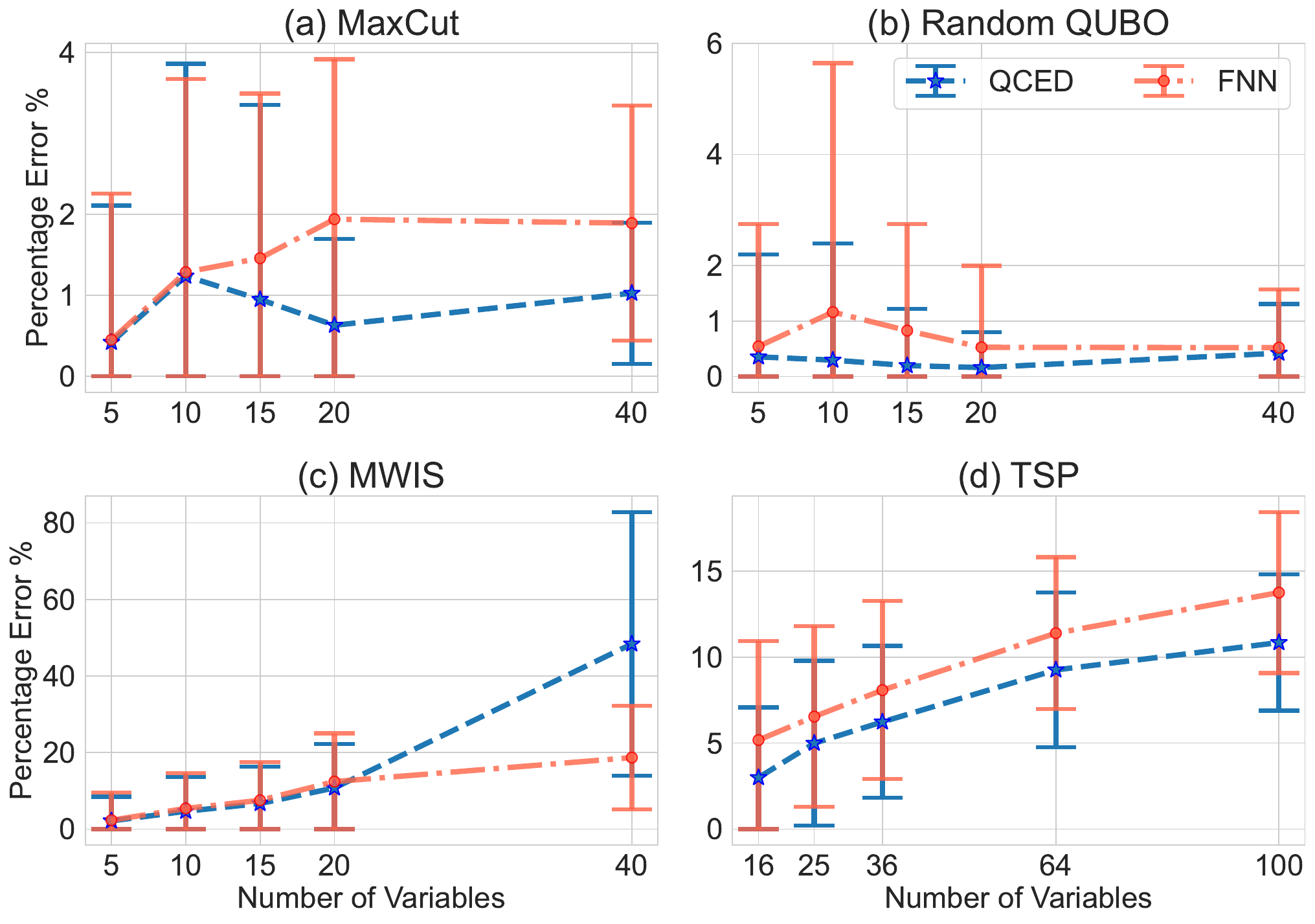}
    \caption{Optimization results of FNN and QCED in 500 iterations of training over 100 samples of MaxCut, random QUBO, MWIS, and TSP, each with varying numbers of variables. The error bars represent one standard deviation of the percentage error calculated across all the samples in the dataset.}
    \label{fig:num_variables_err}
\end{figure}

Next, we examined the scalability of QCED and FNN by analyzing the mean percentage errors across various problem sizes: 5-, 10-, 15-, 20-, and 40-variable MaxCut, random QUBO, and MWIS, as well as 4-, 5-, 6-, 8-, and 10-city TSP problems. The results are shown in Fig.\ref{fig:num_variables_err}, where mean errors are marked by stars and circles, and the error bars represent one standard deviation. Again, the statistics were derived from 100 instances generated for each problem size, while the solvers' architectures were kept consistent as described earlier (i.e., for 40-variable problems, QCED and FNN had 1808 and 264 neurons, respectively). From the plots of  MaxCut, random QUBO, and TSP, a slight advantage of QCED over FNN is noticeable,  although the gap narrows as the problem size approaches 40 variables for MaxCut and random QUBO. On the other hand, FNN performs better for MWIS overall, with the error disparity increasing as the number of variables grows. In the case of MaxCut, both QCED and FNN maintain relatively stable performance across different problem sizes. FNN consistently keeps errors below $2\%$, while QCED stays below $1.3\%$. Similarly, QCED and FNN exhibit strong scalability for random QUBO problems. QCED's optimization results remain well under $1\%$, and FNN only matches this performance for the 40-variable instances, indicating that both solvers handle random QUBO problems proficiently and have a high likelihood of acquiring the exact solutions. For MWIS, however, both optimizers fail to deliver satisfactory results, with errors increasing as the problem size grows. Notably, QCED's scalability appears weak in this case, as the error for 40-node MWIS climbs to nearly $50\%$. Finally, in TSP problems, the optimization errors for both solvers rise steadily with problem size, exceeding $10\%$ for the 10-city problems. The scalability for TSP is not convincing based on this trend, which is likely to continue as more cities are considered.\\

Despite the notable performance of QCED in Fig.\ref{fig:result_15_compare} and Fig.\ref{fig:num_variables_err}, we will demonstrate in the next section that the classical component of QCED actually plays the dominant role of the optimization.

\begin{figure}
    \centering
    \includegraphics[width=\linewidth]{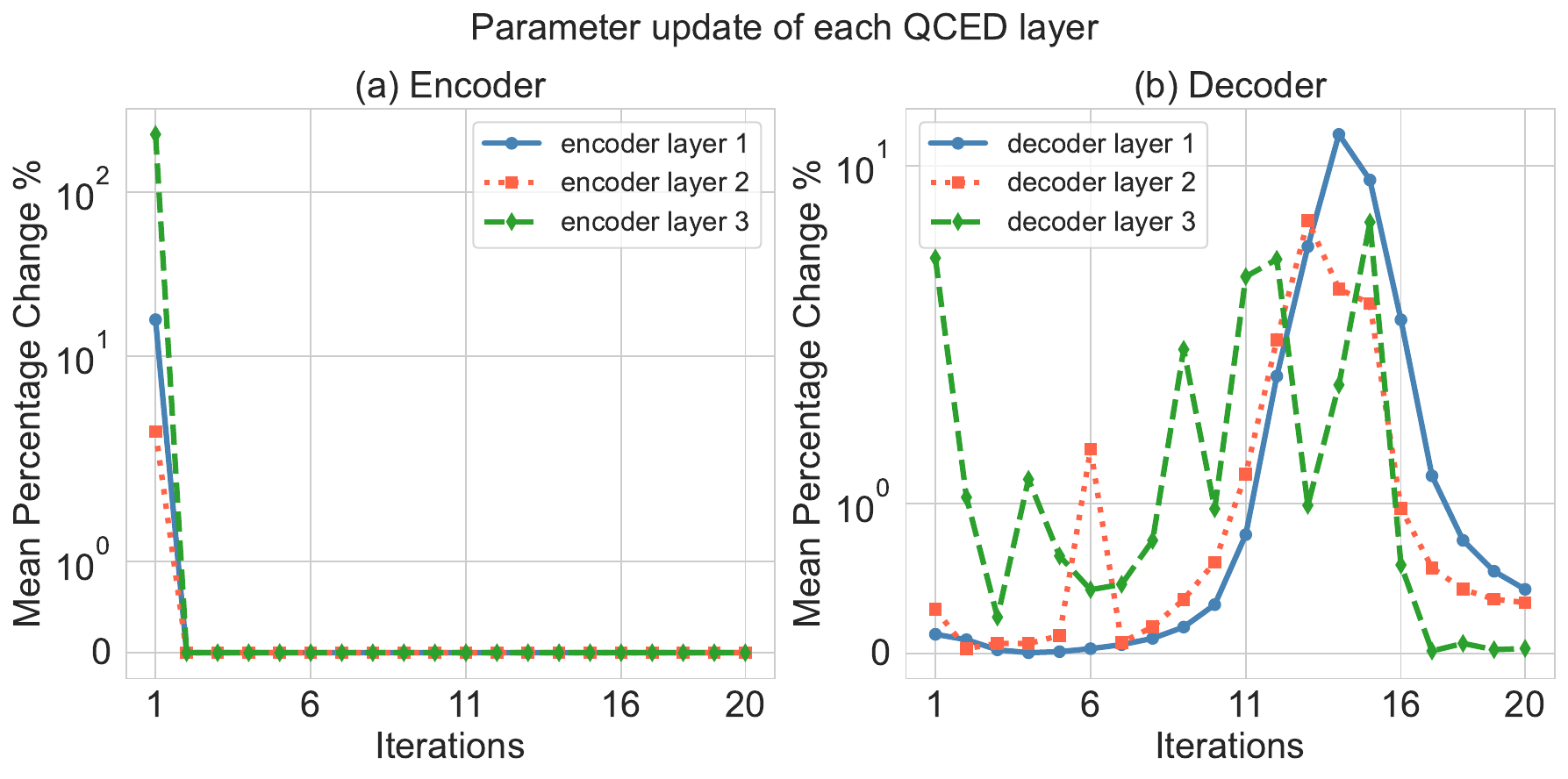}
    \caption{The average percentage change of the parameters in each layer of the encoder and decoder throughout the first 20 iterations of training. The magnitude of the change quantifies the activeness of each layer.}
    \label{fig:QCED_nn_params}
\end{figure}

\subsection{Activeness of the quantum layer in QCED}

Here, we address a fundamental question regarding QCED: What impact does the quantum layer have on the overall optimizer? Does the encoder FNN encode the input QUBO effectively? To answer these questions, we investigated the activeness of neurons in QCED and provided some insights.\\

Since the laser parameters of the annealer are fully encoded by the encoder FNN, the activeness of the quantum layer along with its subsequent outputs are directly tied to the parameter update of the encoder. Consequently, we analyzed how the parameters of the encoder and decoder in QCED were updated over the first 20 iterations of training. At each iteration, we calculated the percentage change in each parameter relative to the previous iteration and averaged these changes to quantify the overall activeness of each layer. From Fig.\ref{fig:QCED_nn_params}, it is evident that all encoder layers remained essentially inactive throughout the process, while the decoder layers were relatively active, with their activity gradually attenuating toward the end. This observation suggests that the components preceding the decoder are utterly deactivated and the quantum layer effectively insulates the backpropagation from reaching the encoder, leaving the decoder as the only functional part of QCED. \\

Given this finding, we anticipated achieving similar optimization performance without the hybrid architecture by utilizing a simple FNN with trainable inputs. Therefore, we extracted the neural network from QCED's decoder to construct the FNN solver for QUBO problems. As previously analyzed, FNN’s performance was comparable to QCED’s on 40-variable problems, with potential for improvement by increasing pre-sampling rounds. Moreover, FNN's lower model complexity and shorter computation time further highlight its advantages over QCED. In conclusion, these observations are strongly in favor of choosing FNN over QCED, as our attempt to integrate a quantum activation function into the neural network provided no tangible benefit in addressing the encoding challenge or significantly enhancing optimization performance. As a result, it is sufficient to continue further benchmarks without QCED.

\subsection{Benchmarking FNN against other optimizers}

\begin{table}[htbp]  
\begin{center}
\adjustbox{max width=\textwidth}{   

\begin{tabular}{|c | c | c || c | c || c | c || c | c|} 
 \hline
 & \multicolumn{2}{c}{MaxCut} & \multicolumn{2}{c}{Random QUBO} &
 \multicolumn{2}{c}{MWIS} & \multicolumn{2}{c|}{TSP} \\
 \hline
 Optimizer & [\textbf{80}] & [\textbf{200}] & [\textbf{80}] & [\textbf{200}] & [\textbf{80}] & [\textbf{200}] & [\textbf{100}] & [\textbf{196}] \\ 
 \hline\hline
 Gurobi & -4918.04 & -27308.26 & -1682.73 & -3934.86 &\cellcolor{yellow} -68.51 & \cellcolor{yellow}-98.50 &\cellcolor{yellow} -19.14 &\cellcolor{yellow} -26.98 \\ 
 \hline
 FNN & -4906.38 & -27530.78 & -1684.35 & -6631.17 & -47.07 & -42.82 & -17.06 & -22.31 \\
 \hline
 FNN+post-processing & \cellcolor{lime} -4929.13 & \cellcolor{lime} -28056.96 & \cellcolor{lime} -1692.32 & \cellcolor{lime}-6768.17 & -59.16 & -66.86 & -17.58 &-25.12\\
 \hline
 Fujitsu emulator & -4906.4 & -28032.9 & -468.61 & -456.79 & 7451.40 & 12603.4 & 280.51 & 303.02 \\ 
 \hline
 D-Wave Advantage & -4496.41 & --- & -1691.74 & --- & 173.30 & --- & -7.73 & ---\\ 
 \hline
 
\end{tabular}
}
\caption{Benchmark of large QUBO problems of varying sizes (the number of variables is denoted in bold) using the Gurobi optimizer, D-Wave Advantage quantum annealer, Fujitsu digital annealer emulator, and FNN. The table records the average QUBO cost values over 100 samples. A lower cost value indicates better performance, as minimization is the main objective.}
\label{table:benchmark}
\end{center}
\end{table}

As shown in Table \ref{table:benchmark}, we benchmarked larger QUBO instances to thoroughly evaluate the potential of the FNN optimizer. Specifically, we tested FNN’s performance on 80- and 200-variable QUBO problems (100- and 196-variable for TSP) by comparing the average QUBO cost values with those obtained by the Gurobi optimizer \cite{gurobi}, the D-Wave Advantage quantum annealer \cite{mcgeoch2020d}, and the Fujitsu digital annealer emulator \cite{matsubara2020digital}. These solvers are among the most advanced in the realm of either quantum or classical optimization in terms of solving large-scale QUBO problems. In these benchmarks, the FNN optimizer was configured with ten neural layers instead of seven in response to the larger problem sizes. A simple post-processing step inspired by classical simulated annealing was also applied to further enhance accuracy, as detailed in \ref{section:postprocessing}. For the Gurobi optimizer, a time limit of 100 seconds was imposed on the solving of each QUBO instance. As for the D-Wave Advantage quantum annealer, a default annealing time of 20 $\mu s$ was applied (excluding additional processing time such as variable encoding and readout), and the best solutions were selected from 100 rounds of annealing. In the Fujitsu emulator, we set the number of iterations per run to $N^{2}$ for an N-variable problem and conducted four runs per instance. Further details on the algorithms behind these solvers are provided in \ref{section:solvers}.\\

According to the average cost values in Table \ref{table:benchmark}, Gurobi excelled in MWIS and TSP problems, consistently obtaining exact solutions for the majority of instances within the time limit, while also delivering high-quality approximate solutions for most other problem types and sizes, as highlighted in yellow. However, it encountered a bottleneck when dealing with 200-variable random QUBOs. For the FNN optimizer, the unprocessed results already demonstrated relatively high accuracy, achieving 99.32\% for 80-variable MaxCut problems and 99.20\% for random QUBO problems. This impressive performance was further augmented by incorporating the post-processing steps (results marked in green in Table \ref{table:benchmark}), which improved the accuracy to 99.78\% and 99.66\% for 80-variable problems, while nudging 200-variable instances by approximately 2\% closer to the optima. These enhancements establish FNN as the best solver for MaxCut and random QUBOs. Despite this, the performance on MWIS was somewhat unsatisfactory, even after post-processing, although it still outperformed the D-Wave annealer and Fujitsu emulator by a wide margin. As for the Fujitsu emulator, it exhibited great scalability for MaxCut problems, but was significantly outperformed by other solvers in other categories. Lastly, the D-Wave quantum annealer managed to optimize 80-variable random QUBOs effectively with accuracy almost identical to that of the post-processed FNN. However, it struggled with other types of problems and was disqualified from the competition for 200-variable instances due to the unsuccessful embedding of variables onto the hardware topology.\\

In summary, FNN albeit not the best solver for every QUBO problem, demonstrates exceptional scalability in MaxCut and random QUBO problems, given its modest resource requirements and model complexity. This advantage will be further investigated in the next subsection, where we analyze the computation time comparison between FNN and Gurobi. 

\begin{figure}
    \centering
    \includegraphics[width=\linewidth]{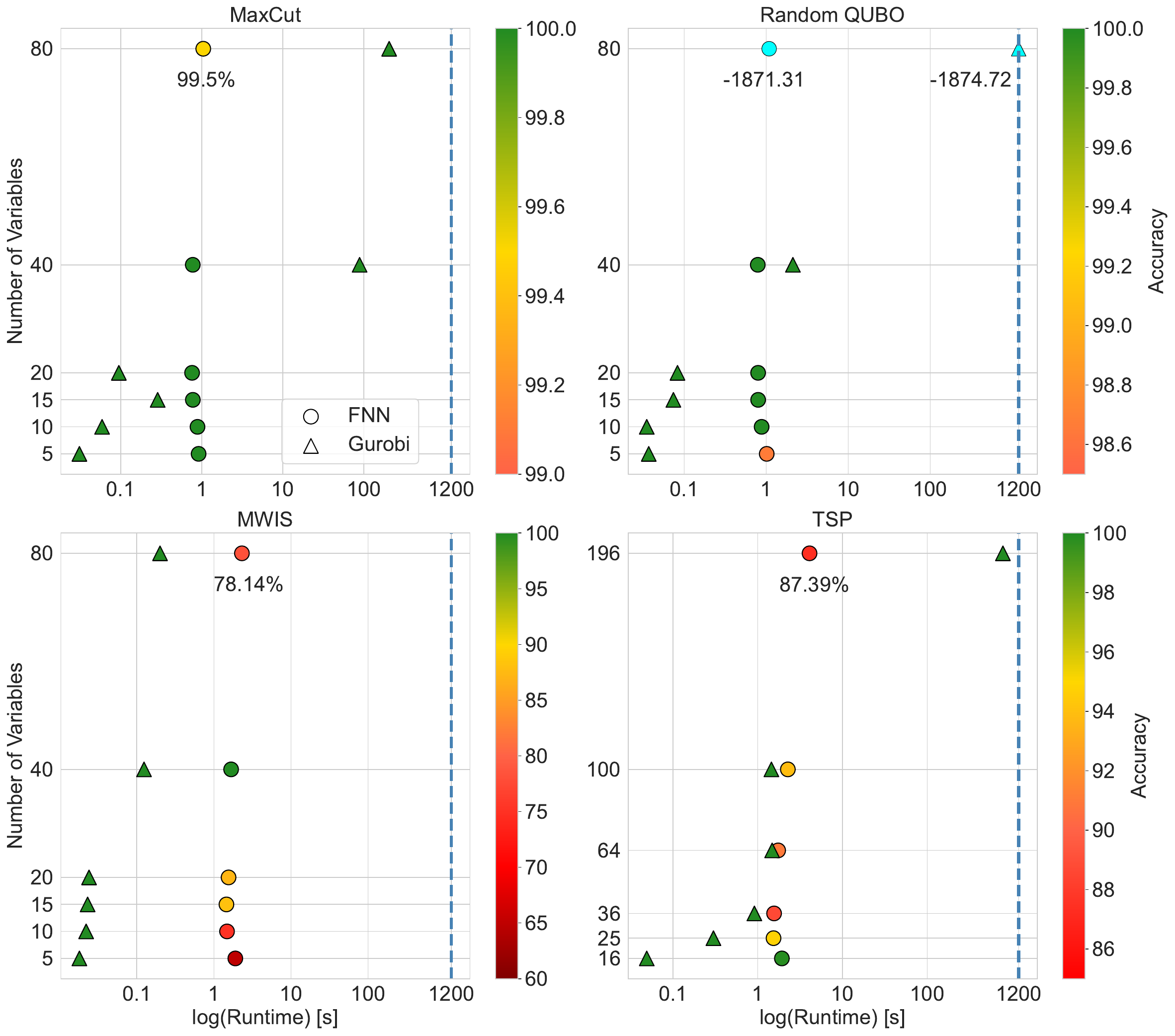}
    \caption{Runtime comparison between FNN and Gurobi across various problem sizes. Marker colors indicate optimization accuracy, while marker shapes differentiate applied solvers (FNN is represented by circles, and Gurobi by triangles). A 1200-second time limit was imposed on Gurobi, with exact solutions provided when the solver converged within this limit. If Gurobi did not converge, the best approximate solution found within the time limit is shown in cyan.}
    \label{fig:runtime}
\end{figure}

\subsection{Computation time of FNN vs. Gurobi}

Apart from optimization accuracy, computation time is another crucial factor in evaluating a solver’s performance. We conducted a runtime benchmark between the FNN and Gurobi optimizers, as displayed in Fig.\ref{fig:runtime}. Both solvers were executed on the same local machine with 8 CPU threads. To prevent Gurobi from performing an exhaustive search, a time limit of 1200 seconds was imposed. Solutions returned by Gurobi within this limit are exact up to numerical precision, while the best-found approximate solution was provided if convergence did not occur within the time limit (indicated in cyan). For FNN, we maintained the previously defined architecture: seven layers for MaxCut, random QUBO, and MWIS problems with 40 or fewer variables (64 or fewer variables in the case of TSP), and ten layers for problems with 80 variables (100 and 196 variables for TSP). MaxCut and random QUBOs were trained for 100 iterations, while MWIS and TSP were trained for 500 iterations. The runtime results show that FNN consistently completes in under 4.1 seconds for MWIS and TSP, and in under 1.1 seconds for MaxCut and random QUBOs, regardless of the problem size. In contrast, Gurobi’s runtime is generally under 1 second for smaller problems, but it increases significantly—by 2 to 3 orders of magnitude—as the number of variables doubles. For the 80-variable MaxCut and random QUBO problems, FNN achieved an accuracy of 99.5\% and a cost value almost identical to Gurobi's, while using only approximately 1/200 and 1/1200 of the time, respectively. However, for MWIS, Gurobi demonstrated better scalability, solving the 80-variable instance in approximately 0.2 seconds, whereas FNN achieved only 78.14\% accuracy in 2.3 seconds. These observations highlight FNN’s exceptional runtime advantage over Gurobi for larger QUBO problems, except for MWIS. Its consistently short computation times and solid optimization accuracy make FNN a powerful and efficient heuristic for fast QUBO optimization.

\section{Conclusions and Outlook}
\label{Conclusion}

 We successfully designed a highly resource-efficient FNN optimizer capable of handling arbitrary QUBO problems using no training data. This optimizer demonstrated impressive performance, particularly with high-degree MaxCut problems and dense random QUBOs, achieving accuracy over 99\% in less than 1.1 seconds, while also providing reasonably good approximate solutions to TSP problems. Regarding the question of enhancing the FNN optimizer by incorporating a quantum annealer as a quantum activation function, we discovered, through the comparative analysis and neuron activity examination, that such a combination offers no benefit to the classical FNN mainly due to the deactivation of the quantum layer. Given the evidence, we proceeded with the investigation of the FNN solver by benchmarking both the optimization performance and the computation time against state-of-the-art optimizers such as Gurobi. While the achieved accuracy was comparable, FNN exhibited a significant runtime advantage over Gurobi for larger problems, outperforming it by 2 to 3 orders of magnitude. This feature can be capitalized in aiding \textit{large-scale} and \textit{real-time} optimization problems like smart grid optimization \cite{ullah2022quantum}, traffic flow optimization \cite{neukart2017traffic}, portfolio optimization in high-frequency trading \cite{tatsumura2023real}, telecommunications \cite{ramirez2020integrating}, and anomaly detection for cybersecurity \cite{wang2022integrating}.\\

Despite the inactivity of the quantum layer, exploring other potential uses of QCED is still worthwhile, where the quantum activation function plays a more significant role. Interestingly, Rydberg atom arrays are the quantum equivalence of Recurrent Neural Networks (RNNs) in the limit of diagonal Hamiltonians, as highlighted by \cite{bravo2022quantum}. This implies that interpreting QCED as an RNN embedded within FNNs may open up new applications beyond QUBO optimization. However, further research is needed to better understand the interaction between quantum and classical components in hybrid networks, and to identify conditions under which quantum layers remain active. In terms of improving the FNN solver, a promising direction for future research could be introducing Bayesian-based approaches in the optimization routine \cite{wu2019hyperparameter,victoria2021automatic,snoek2015scalable,jospin2022hands,mukherjee2020preparation}. Lastly, the relative performance across different types of QUBO problems seems to suggest that QUBO matrices with high levels of randomness such as random QUBO and high-degree weighted MaxCut, tend to be easier for FNN to address, whereas more structured matrices like MWIS pose greater challenges not only for FNN but also for other heuristics such as quantum and digital annealers. In light of this, we offer some perspectives on the topic in \ref{app:hardness_qubo}; nevertheless, the inherent complexity of QUBO problems still requires further study.

\ack

This work is funded by the German Federal Ministry of Education and
Research within the funding program “Quantum Technologies - from basic
research to market” under Contract No. 13N16138.

\section*{Code Availability}

The data and code that support the findings of this study are openly available at the following GitHub repository: \href{https://github.com/jordan0809/Classical-and-Quantum-Activated-Feedforward-Neural-Networks}{Classical and Quantum-Activated FNNs}

\section*{Conflict of Interest}
Carsten Blank, CEO and Co-Founder of Data Cybernetics SSC GmbH, and the other authors declare no conflicts of interest regarding the publication of this paper.

\appendix

\section{\label{app:gradient_backpropogation}Gradients Evaluation \& Backpropagation}

As part of a hybrid neural network, optimizing the laser parameters in the Rydberg annealer is necessary for the utility of QCED. In this work, we opted for the gradient descent approach to train QCED, which demands the evaluation of gradients and a coherent backpropagation through the quantum and classical layers of the optimizer \cite{hecht1992theory,werbos1990backpropagation}. To calculate the gradients of an analog quantum system, the finite-difference method is the most efficient option among others. Therefore, the gradient of $\Psi_{z}^{j}(\bm{\theta})$ (Eq.(\ref{eq:expectation})) with respect to a laser parameter $\theta_{m} \in \bm{\theta} =(\Omega_{i},\Delta_{k}(0),s_{k})$ is evaluated as follows:

\begin{equation}\label{eq:finite_difference}
\pdv{\Psi_{z}^{j}(\bm{\theta})}{\theta_{m}} = \frac{\Psi_{z}^{j}(\bm{\theta}+\epsilon \bm{e}_{m})-\Psi_{z}^{j}(\bm{\theta}-\epsilon \bm{e}_{m})}{2\epsilon} \ ,
\end{equation}
where $\bm{e}_{m}$ is a unit vector with all elements equal to zero except for the one associated with the $m^{th}$ parameter. For the evaluation of a gradient, two rounds of simulations are required to obtain expectation values for different laser configurations. As a consequence, the number of simulations scales linearly with the number of laser parameters. To avoid the tedious training process for larger quantum systems, a global Rabi frequency can be applied instead of local ones, which also reduces the difficulty in terms of experimental control.\\

The gradients of the Rydberg annealer are constituents of the backpropagation and can be combined with those of other classical parameters via the chain rule to compute the gradient for the final loss function. The product of quantum and classical gradients forms a cohesive backward pass that can reach any desired parameter in the hybrid network, enabling an unimpeded workflow of gradient descent. Consider a simple combination of a Rydberg annealer with $q$ qubits and a layer of classical neural network with $P$ neurons, the gradient of the loss function $\mathcal{L}$ w.r.t. any laser parameter $\bm{\theta}_{m}$ can be computed as:

\begin{equation}
    \pdv{\mathcal{L}}{\theta_{m}} = \sum_{j=1}^{q}\sum_{p=1}^{P}\pdv{\mathcal{L}}{\sigma(g_{p})}\pdv{\sigma(g_{p})}{g_{p}}\pdv{g_{p}}{\Psi_{z}^{j}(\bm{\theta})}\pdv{\Psi_{z}^{j}(\bm{\theta})}{\bm{\theta}_{m}} \ ,
\end{equation}\\
where $g_{p} \equiv g_{p}(\Psi_{z})$ is the output of the $p^{th}$ neuron stemming from some linear transformation of $\Psi_{z}(\bm{\theta})$ from the quantum layer and $\sigma(g_{p})$ is the activation function applied to this output. The process of backpropagation can be further extended to networks with arbitrary orders of quantum and classical layers as long as the parameters are connected and the differentiation of the quantum layer is possible.\\

Aside from the finite-difference method, an approximated parameter-shift rule can also be applied to analog quantum systems given a sufficiently short evolution time. To simplify the derivation of the approximated parameter-shift rule, we consider a Rydberg Hamiltonian whose Rabi frequencies and detunings are both localized and adopt a laser schedule of a discrete step function for every local $\Omega_{j}(t)$ and $\Delta_{j}(t)$ without loss of generality since we can represent any function given enough steps. The Rydberg Hamiltonian is then time-independent within each equally long time step $t_{i} \in (i\Delta t,(i+1)\Delta t]$ and is parametrized by the laser parameters $\Omega_{j}(t_{i})$ and $\delta_{j}(t_{i})$ for every qubit and every step. Here we use the shifted detuning $\delta_{j} = \Delta_{j}-\sum_{k \neq j}V_{jk}/2\hbar$ for convenience. The Hamiltonian $H_{Ryd}(t_{i})$ for the $i^{th}$ time step reads:

\begin{equation}
H_{Ryd}(t_{i}) = \frac{\hbar}{2} \sum_{j}\Omega_{j}(t_{i})\sigma_{j}^{x}- \frac{\hbar}{2} \sum_{j}\delta_{j}(t_{i})\sigma_{j}^{z}+ \frac{\hbar}{2}\sum_{<j,k>}V_{jk}^{\prime}\sigma_{j}^{z}\sigma_{k}^{z} \ ,
\end{equation}
which can be derived from Eq.(\ref{eq:H_ryd2}) by taking $n_{j} = (1+\sigma_{j}^{z})/2$ and $V_{jk}^{\prime}=V_{jk}/2\hbar$. Given this stepwise Rydberg Hamiltonian, we can decompose the corresponding unitary time evolution into a sequence of sub-unitaries associated with time-independent Hamiltonians over some discretized time intervals.

\begin{equation}
    U_{Ryd}(\bm{\theta}) = \prod_{i=0}^{N-1} U(t_{i};\bm{\theta}(t_{i})) \quad , \quad 
    U(t_{i};\bm{\theta}(t_{i})) = \exp[-\frac{i}{\hbar}H_{Ryd}(t_{i})\cdot \Delta t] \ ,
\end{equation}
where $\bm{\theta}=(\Omega_{j},\delta_{j})$ represents the laser parameters. These sub-unitaries $U(t_{i};\bm{\theta}(t_{i}))$ can be further decomposed with the help of Trotterization \cite{hatano2005finding,suzuki1976generalized}. Using the Suzuki-Trotter formula, $U(t_{i};\bm{\theta}(t_{i}))$ is approximated as follows:

\begin{equation}
\begin{split}
    U(t_{i};\bm{\theta}(t_{i})) &= \Bigg[\exp(\frac{-i\Delta t}{4n}\sum_{j}\Omega_{j}(t_{i}) \sigma_{j}^{x})\exp(\frac{i\Delta t}{2n}\sum_{j}\delta_{j}(t_{i})\sigma_{j}^{z}) \\ & \quad \cdot \exp(\frac{-i\Delta t}{2n}\sum_{<j,k>}V_{jk}^{\prime}\sigma_{j}^{z}\sigma_{k}^{z})\exp(\frac{-i\Delta t}{4n}\sum_{j}\Omega_{j}(t_{i}) \sigma_{j}^{x})\Bigg]^{n}+\mathcal{O}(\frac{(\Lambda \Delta t)^{3}}{n^{2}}) \\ &= \Bigg[\prod_{j}\exp(\frac{-i\Delta t}{4n}\Omega_{j}(t_{i}) \sigma_{j}^{x})\prod_{j}\exp(\frac{i\Delta t}{2n}\delta_{j}(t_{i})\sigma_{j}^{z}) \\ & \quad \cdot \prod_{<j,k>}\exp(\frac{-i\Delta t}{2n}V_{jk}^{\prime}\sigma_{j}^{z}\sigma_{k}^{z})\prod_{j}\exp(\frac{-i\Delta t}{4n}\Omega_{j}(t_{i}) \sigma_{j}^{x})\Bigg]^{n}+\mathcal{O}(\frac{(\Lambda \Delta t)^{3}}{n^{2}})\\
    &= \Bigg[  \prod_{j}R_{x_{j}}(\frac{\Delta t}{2n}\Omega_{j}(t_{i}))\prod_{j}R_{z_{j}}(\frac{-\Delta t}{n}\delta_{j}(t_{i})) \\  & \quad \cdot \prod_{<j,k>}R_{z_{j}z_{k}}(\frac{\Delta t}{n}V_{jk}^{\prime})\prod_{j}R_{x_{j}}(\frac{\Delta t}{2n}\Omega_{j}(t_{i}))\Bigg]^{n}+\mathcal{O}(\frac{(\Lambda \Delta t)^{3}}{n^{2}}) \ ,
\end{split}
\end{equation}
where $n$ is the Trotter number and $\Lambda=\max \{\norm{H_{x}},\norm{H_{z}},\norm{H_{zz}}\}$ is the upper bound of the spectral norm with $H_{x} = \frac{1}{2}\sum_{j}\Omega_{j}(t_{i})\sigma_{j}^{x}$, $H_{z}=\frac{1}{2}\sum_{j}\delta_{j}(t_{i})\sigma_{j}^{z}$, and $H_{zz} =\frac{1}{2}\sum_{<j,k>}V_{jk}^{\prime}\sigma_{j}^{z}\sigma_{k}^{z}$. As a result, the full unitary time evolution is 

\begin{equation}\label{eq:Trotter}
\begin{split}
    U_{Ryd}(\bm{\theta}) &=\prod_{i=0}^{N-1}\Bigg[  \prod_{j}R_{x_{j}}(\frac{\Delta t}{2n}\Omega_{j}(t_{i}))\prod_{j}R_{z_{j}}(\frac{-\Delta t}{n}\delta_{j}(t_{i})) \\ & \quad \quad \quad \cdot \prod_{<j,k>}R_{z_{j}z_{k}}(\frac{\Delta t}{n}V_{jk}^{\prime})\prod_{j}R_{x_{j}}(\frac{\Delta t}{2n}\Omega_{j}(t_{i}))\Bigg]^{n}+\mathcal{O}(\frac{N(\Lambda \Delta t)^{3}}{n^{2}})
\end{split}
\end{equation}
This way, the parameter-shift rule for repetitive parameters \cite{schuld2019evaluating,crooks2019gradients,wierichs2022general,banchi2021measuring} can be introduced to $\Psi_{z}^{j}(\bm{\theta})$ to obtain approximated gradients :

\begin{figure}
    \centering
    \includegraphics[scale=0.5]{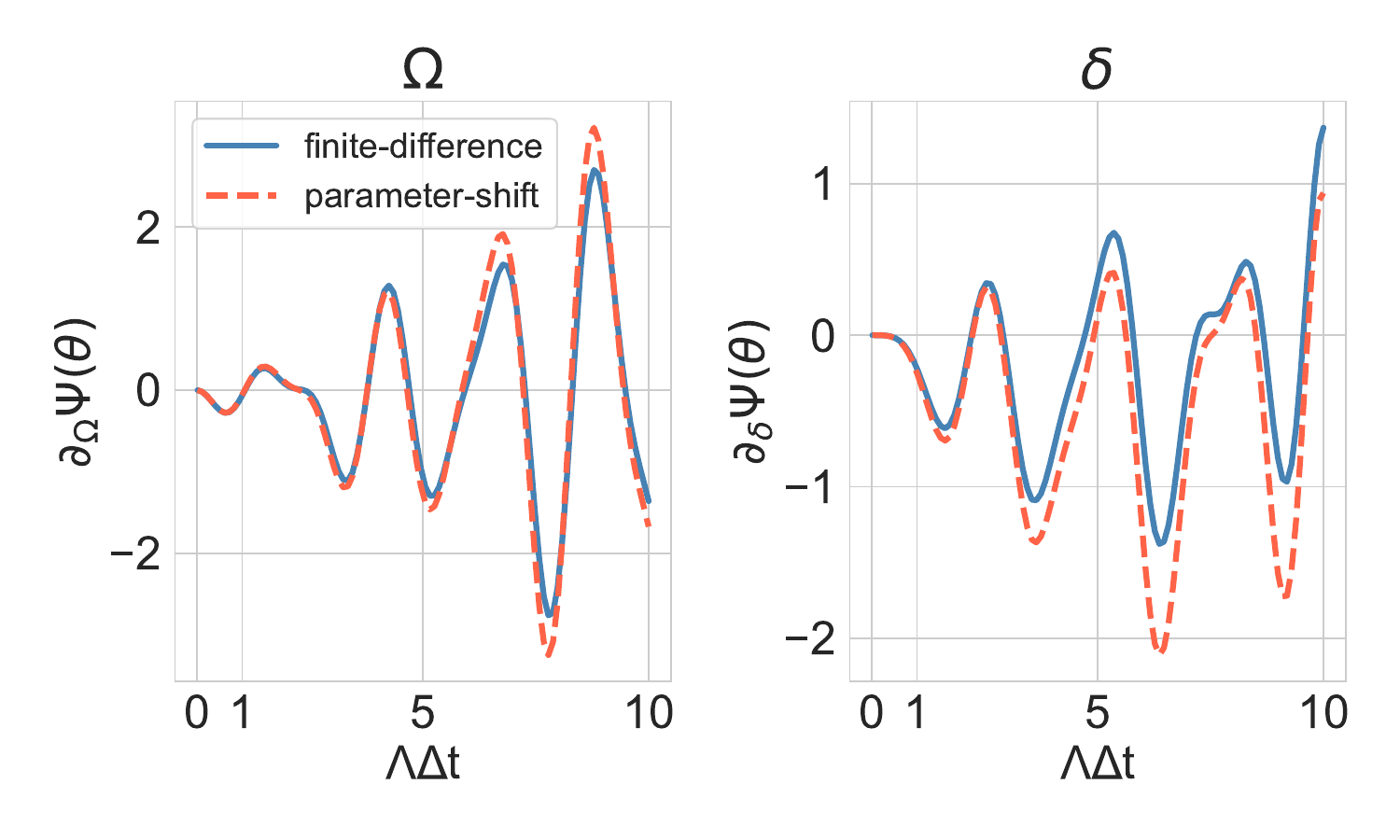}
    \caption{Gradients evaluated with finite-difference method and the approximated parameter-shift rule for different lengths of a discretized time step. We chose $n=2$ for the parameter-shift rule. The macroscopic trends of both differentiation methods are identical, except that the gradients gradually diverge towards larger $\Lambda \Delta t$, as the error of the approximated parameter-shift rule increases with the duration.}
    \label{fig:parameter_shift}
\end{figure}

\begin{equation}
\begin{split}
    \pdv{\Psi_{z}(\bm{\theta})}{\Omega_{j}(t_{i})} &\approx \frac{\Delta t}{4n}\sum_{l=1,n+1}[\Psi_{z}(\Omega_{j}^{(l)}(t_{i})+\frac{n\pi}{\Delta t})-\Psi_{z}(\Omega_{j}^{(l)}(t_{i})-\frac{n\pi}{\Delta t})]\\ &+\frac{\Delta t}{2n}\sum_{l=2}^{n}[\Psi_{z}(\Omega_{j}^{(l)}(t_{i})+\frac{n\pi}{2\Delta t})-\Psi_{z}(\Omega_{j}^{(l)}(t_{i})-\frac{n\pi}{2\Delta t})]
\end{split}
\end{equation}
\begin{equation}
    \pdv{\Psi_{z}(\bm{\theta})}{\delta_{j}(t_{i})} \approx \frac{\Delta t}{2n}\sum_{l=1}^{n}[\Psi_{z}(\delta_{j}^{(l)}(t_{i})+\frac{n\pi}{2\Delta t})-\Psi_{z}(\delta_{j}^{(l)}(t_{i})-\frac{n\pi}{2\Delta t})] \ ,
\end{equation}
where $l$ labels the order of the repetitive gates in the Trotterization.\\

Unlike the parameter-shift rule for parametrized circuits, the gradient evaluations of the laser parameters for Rydberg annealers are not exact. Besides, the number of simulations required scales with the Trotter number $n$, which in turn determines the accuracy. Therefore, the approximated parameter-shift rule is not as ideal a strategy of gradient evaluation as the finite-difference method for analog systems. Nevertheless, two gradient methods are compared in Fig.\ref{fig:parameter_shift} to demonstrate the validity of our argument. Both strategies exhibit similar trends of gradient values across different time durations $\Lambda \Delta t$, though the gradient values obtained using the parameter-shift rule increase with longer durations, as predicted in Eq.(\ref{eq:Trotter}).\\

\section{QUBO Dataset}
\label{app:QUBO Problems}

In this appendix, we go through different classes of QUBO problems we solved and describe how we generated the dataset for these instances. Exemplary QUBO matrices from our dataset are demonstrated in Fig.\ref{fig:different_matrices}.

\subsection{MWIS instances}

\begin{figure}
    \centering
    \subfigure[]{\includegraphics[width=0.3\textwidth]{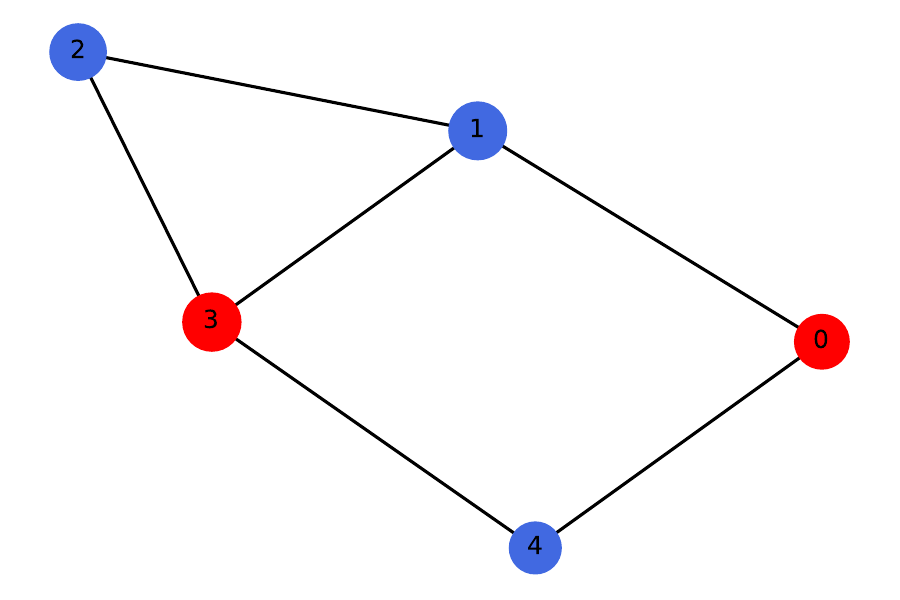}
    \label{fig:5MWIS}}
    \hfill
    \subfigure[]{\includegraphics[width=0.3\textwidth]{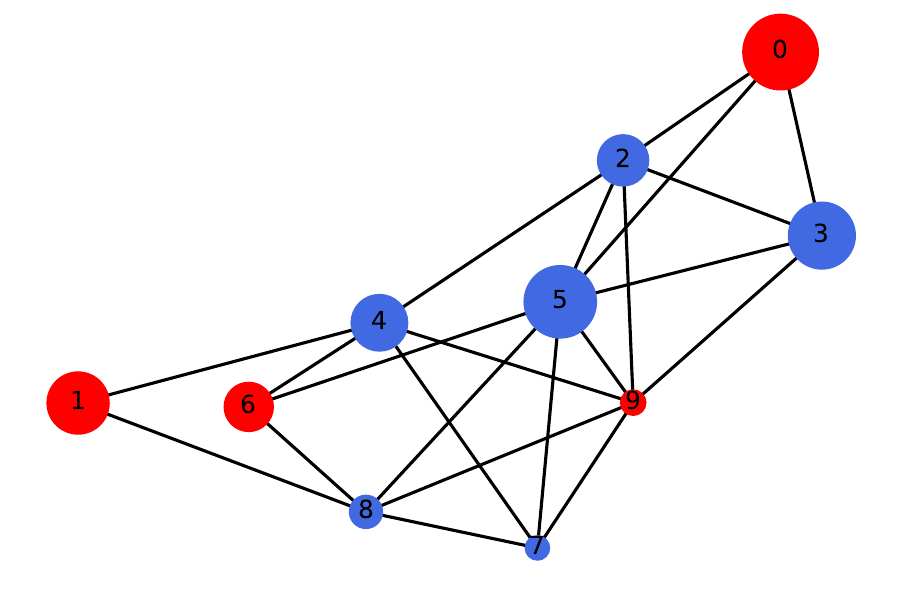}\label{fig:10MWIS}}
    \hfill
    \subfigure[]{\includegraphics[width=0.3\textwidth]{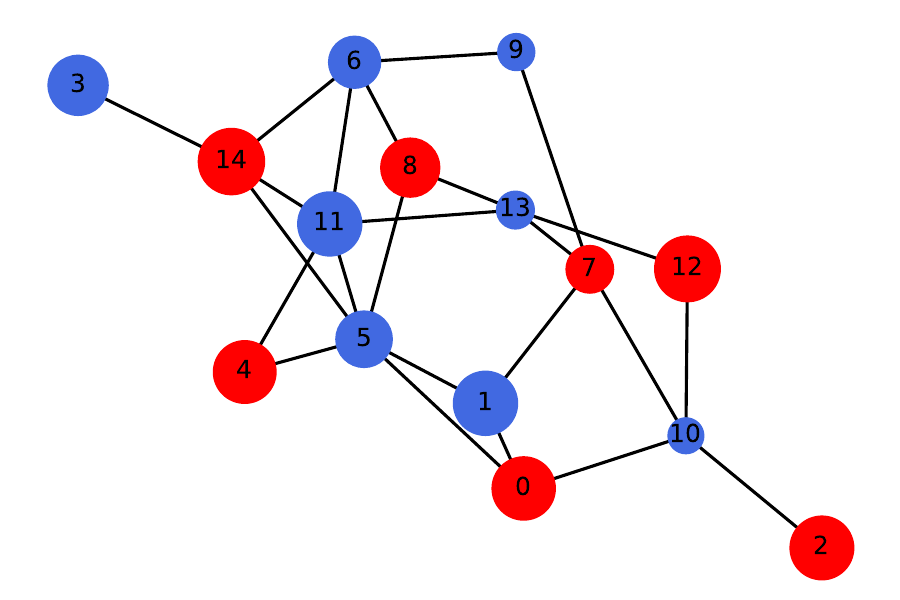}\label{fig:15MWIS}}
\caption{Example graphs with 5, 10, and 15 nodes from the MWIS dataset. The red nodes mark the MWIS of the graph while the size of the nodes corresponds to their weights.}
\label{fig:MWIS_data}
\end{figure}

To generate the dataset of MWIS, we first created random problem graphs with 5, 10, 15, 20, 40, and 80 nodes respectively. For each $n$-node graph, we randomly selected up to $n(n-1)/2$ links from all the possible connections without limiting the degree of each node. Then, we sampled node weights from a uniform distribution ranging from 0 to 10 and assigned them to the nodes. Fig.\ref{fig:MWIS_data} shows an example graph of three different problem sizes from the dataset.\\
\\
The QUBO cost function for MWIS can be formulated as:

\begin{equation}
    x^{T}Qx = -\sum_{i}W_{i}x_{i}+\sum_{i}\sum_{j\in N(i)}Px_{i}x_{j} \ ,
\end{equation}\\
where $W_{i}$ is the weight of the $i^{th}$ node, $P>0$ is the penalty, which we set to 10 for every instance, and $N(i)$ denotes the set of neighbors of node $i$.

\subsection{MaxCut instances}

\begin{figure}
    \centering
    \subfigure[]{\includegraphics[width=0.3\textwidth]{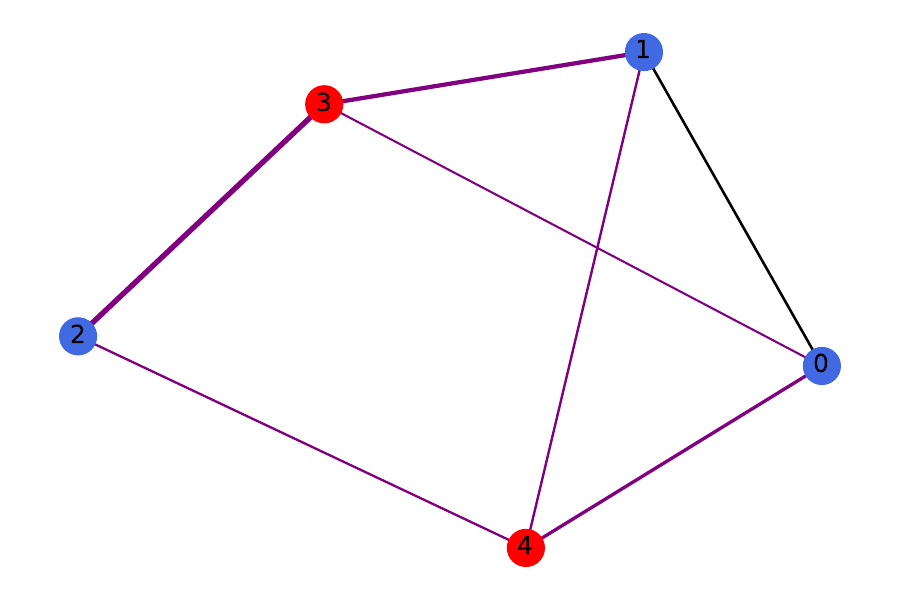}
    \label{fig:5maxcut}}
    \hfill
    \subfigure[]{\includegraphics[width=0.3\textwidth]{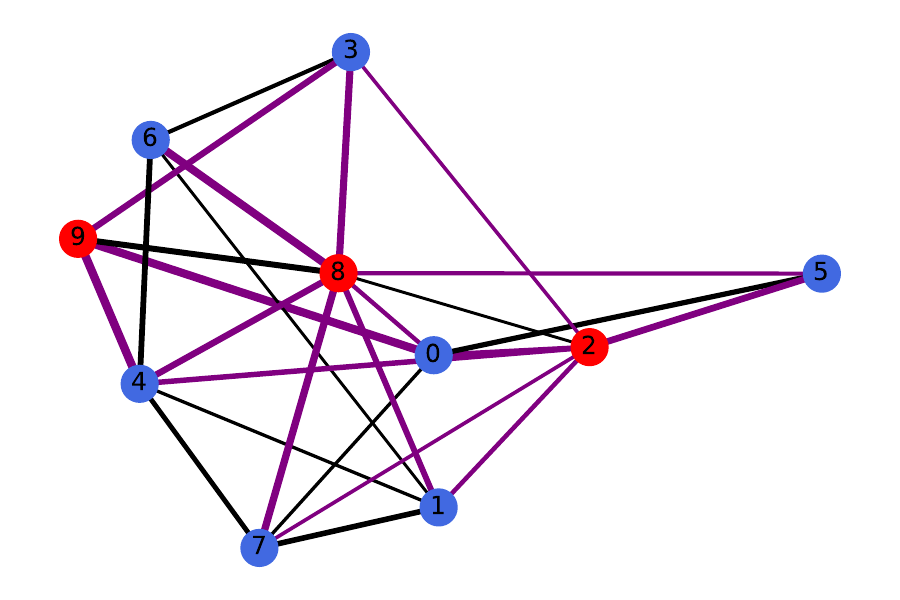}\label{fig:10maxcut}}
    \hfill
    \subfigure[]{\includegraphics[width=0.3\textwidth]{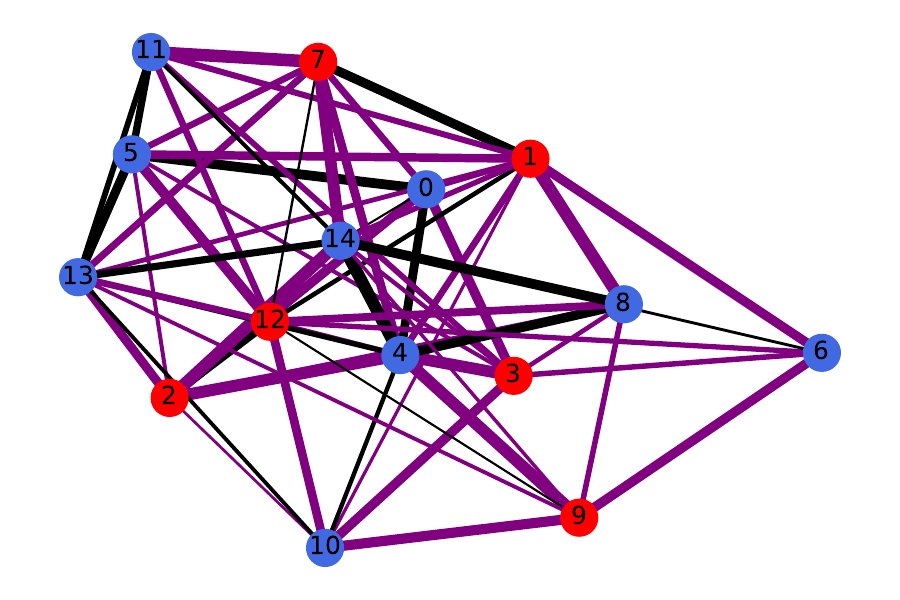}\label{fig:15maxcut}}
\caption{Example graphs from the MaxCut dataset. The blue and red nodes mark two different partitions of the optimal cut. The edge widths correspond to the weights and the cut-through edges are colored in purple.}
\label{fig:maxcut_data}
\end{figure}

MaxCut is perhaps the most well-known QUBO problem and serves as a standard benchmark for optimization algorithms. The objective of MaxCut is to find a single cut through the edges of a graph that partitions the nodes into two sets such that the sum of edges connecting nodes in different partitions is maximized. A more advanced version of MaxCut is the weighted MaxCut, where weights are assigned to the edges, and the sum of the weights of edges crossing the cut is to be maximized. We generated the weighted MaxCut instances by creating graphs of random connections. Then, we assigned random weights to the edges from a uniform distribution ranging from 0 to 10 (Fig.\ref{fig:maxcut_data}).\\
\\
The QUBO cost function for MaxCut can be expressed as:

\begin{equation}
    x^{T}Qx = -\sum_{i}\sum_{j}W_{ij}x_{i}(1-x_{j}) = -\sum_{i}(\sum_{j}W_{ij})x_{i}+\sum_{i}\sum_{j \neq i}W_{ij}x_{i}x_{j} \ ,
\end{equation}\\
where $W_{ij}$ denotes the weight of the edge $E(i,j)$.

\subsection{Random QUBO instances}

By random QUBO instances, we refer to randomly generated QUBO matrices with elements drawn from a uniform distribution between -10 and 10, without reference to any specific optimization problem, in the interest of generality. Technically speaking, MaxCut, MWIS, and TSP problems all fall into this category, and so do other QUBO problems. However, in our dataset, the randomly sampled elements typically result in dense symmetric matrices, unlike the sparsely populated matrices sometimes encountered in MaxCut and MWIS problems, where many off-diagonal elements are zero disconnected nodes. Furthermore, random QUBO instances in our dataset are characterized by a relatively high degree of disorder and an absence of clear repetitive patterns, in contrast to other problem classes we considered. Examples of optimization problems with similar characteristics to our random QUBO matrices include set packing \cite{delorme2004grasp}, set partitioning \cite{balas1976set}, quadratic assignment \cite{lawler1963quadratic}, and multiway number partitioning \cite{korf2009multi}, just to name a few. The core message we aim to convey here is that random QUBO matrices are by no means impractical. On the contrary, they have a broad range of applications extending beyond specific problem types considered in this work.

\subsection{\label{sec:TSP}TSP}

TSP is another renowned yet challenging QUBO problem that attracts much attention in the field of optimization. TSP aims to find the shortest round trip through several locations without missing or re-visiting any of them (except for the starting point). The challenge arises from the permutation nature of the problem and the hard constraints that come with it. The QUBO cost function for TSP can be divided into three parts: minimizing the traveled distance, enforcing the constraint that each location is visited exactly once, and ensuring that only one location is visited at each step. Explicitly, the cost function reads:

\begin{equation}
\begin{split}
    C_{TSP} &= \sum_{ij}W_{ij}\sum_{p}x_{i,p}x_{j,p+1}+A\sum_{p}(1-\sum_{i}x_{i,p})^{2}\\ &+A\sum_{i}(1-\sum_{p}x_{i,p})^{2} \ ,
\end{split}
\end{equation}
where $W_{ij}$ is the distance between the $i^{th}$ and $j^{th}$ locations, $x_{i,p}$ is the binary variable representing whether the $i^{th}$ location is visited at step $p$, and A is a penalty constant. For a TSP with $n$ locations, $n^{2}$ variables are required to keep track of both the locations and their permutation in the tour. To prepare TSP instances, we created random distance matrices and normalized them to the domain of $(0,1]$ for better processing in the neural network. The penalty constant $A$ was chosen as $10\%$ greater than the maximal distance (which is 1 after normalization).

\begin{figure}
    \centering
    
    \subfigure[]
    {\includegraphics[width=0.48\textwidth]{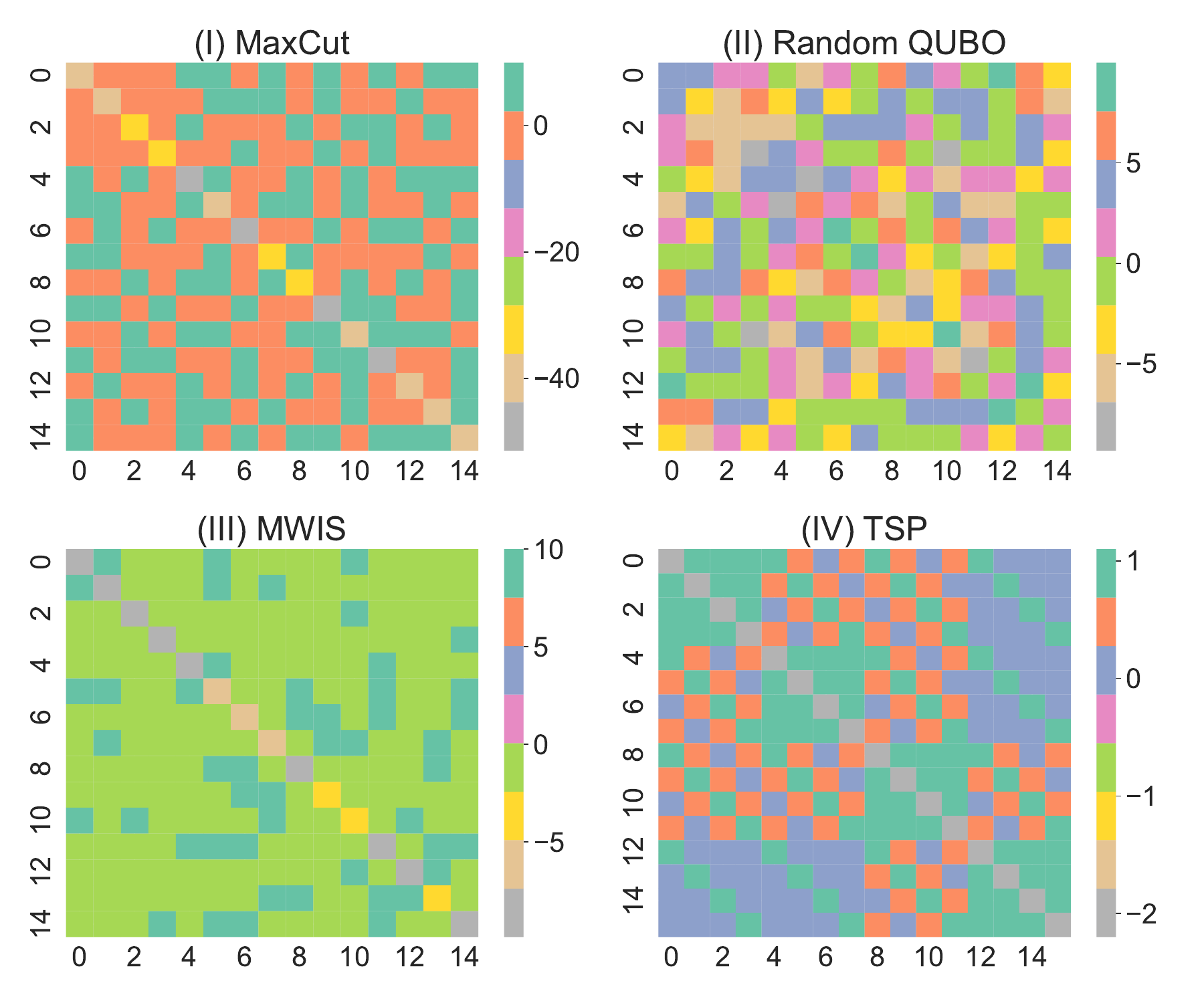}\label{fig:different_matrices}}
    \hfill
    \subfigure[]
    {\includegraphics[width=0.48\textwidth]{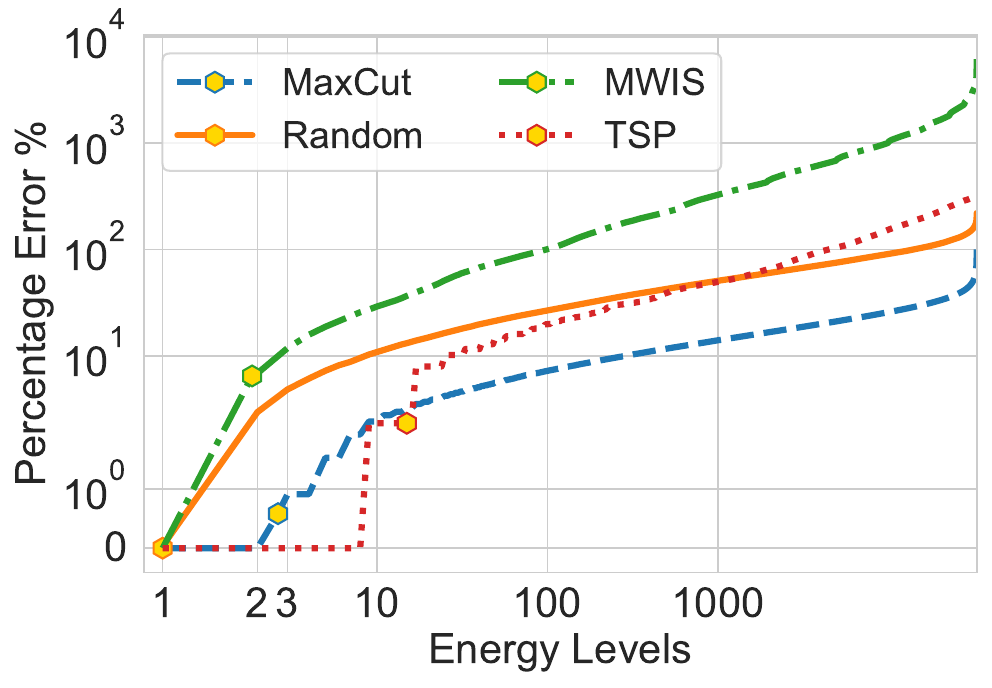}
    \label{fig:energy_level}}
    
\caption{(a) Sample matrices from the dataset of MaxCut, random QUBO, MWIS, and TSP. (b)The average cost errors for different energy levels across all the problem classes we benchmarked. The yellow markers indicate the average optimization results obtained using FNN.}
\label{fig:matrices+levels}
\end{figure}

\section{Hardness of QUBO Problems}
\label{app:hardness_qubo}

There are several factors that can contribute to the hardness of a QUBO problem besides the number of variables. For instance, the number of interaction terms directly influences how easily the problem can be encoded onto quantum hardware, as finding an embedding that accurately reflects the interactions becomes more challenging when more qubits are coupled. Consequently, problems with dense QUBO matrices are generally considered more complex than those with sparser matrices. Additionally, the energy landscape of the optimization task plays a role in determining the difficulty of finding the optimal solution. Narrow spectral gaps in the energy landscape, for example, are widely believed to increase the complexity of the problem. However, opposite trends have been observed in empirical studies of machine learning, where a negative correlation between gap size and difficulty was found \cite{gerlach2024investigating}. Moreover, frequent distributions of low-energy excited states with short Hamming distances from the ground state often hinder quantum annealers' ability to solve the problem successfully \cite{mehta2022hardness}. Beyond these commonly discussed factors, we offer a different perspective from which we can analyze the hardness of QUBO instances.\\

Throughout this work, we have adopted the percentage error of the QUBO cost function as the benchmark for evaluating solution quality. The percentage error is defined as the difference between the optimal and the obtained cost values, divided by the optimal cost value. However, this evaluation criterion might create bias due to negligence of the energy levels. For instance, a QUBO cost function with narrow gaps between energy levels often results in a lower percentage error in terms of energy loss, compared to a cost function with wider gaps. In other words, for two solutions occupying the same energy levels in different problems, the associated percentage errors could be vastly different. Therefore, inspecting both the energy level and the corresponding cost error provides a more comprehensive understanding of the optimization result and insights into the difficulty of the QUBO instance. To illustrate this, we plotted the relationship between energy levels and their corresponding average cost errors over 100 samples in Fig.\ref{fig:energy_level}. The diagram elucidates this benchmark bias more clearly with the optimization results of FNN marked in yellow. As displayed, the solution for random QUBO exhibits a lower percentage error and also resides at a lower energy level compared to the solutions for all other QUBO classes, indicating that FNN unarguably performs better on random QUBO by both measures. However, when comparing MaxCut (or TSP) with MWIS, we notice that the percentage error for MaxCut (or TSP) is lower than for MWIS, despite the solution populating a higher energy level. This contrast perfectly highlights the relative hardness between these different problem types. In conclusion, the interpretation of a benchmark result is contingent on the criteria used to evaluate solution quality.

\section{State-of-the-art QUBO Optimizers}
\label{section:solvers}

In this appendix, we briefly introduce the algorithms used by the optimizers in our benchmarking of large QUBO problems. Since the implementation details of these solvers are proprietary, we will only offer a general and high-level overview of the methodologies.

\subsection{Gurobi Optimizer}
\label{section:Gurobi}

The Gurobi optimizer is a powerful optimizer for various mathematical programming models. In particular, Gurobi prides itself on the capability of solving continuous linear programming problems (LP) \cite{dantzig2002linear,ignizio1994linear}. That being said, Gurobi is also capable of solving a variety of programming models other than LP by relaxation of discrete variables and introduction of constraints to linearize quadratic terms. The optimizer supports multiple variable types, including continuous, integer, binary, semi-continuous, and semi-integer. In the case of QUBO, a standard procedure is to transform the problem into mixed integer programming (MIP) \cite{wolsey2007mixed,smith2008tutorial} whose objective can be formulated as follows:

\begin{equation}
    \min{c^{T}x} \quad , \quad Ax \leq b \ ,
\end{equation}
where some or all of the variables $x$ are integers. The transformation from QUBO to MIP is enabled by first defining new binary variables $y_{ij}=x_{i}x_{j}$, representing the product of every pair of original variables. This way, the QUBO problem is essentially linearized. Then, some inequality constraints need to be imposed on the newly defined variables to ensure the feasibility of the solution. These constraints are given by:

\begin{equation}\label{eq:mip_constraint}
\begin{split}
    y_{ij} &\leq x_{i} \\
    y_{ij} &\leq x_{j} \\
    y_{ij} &\geq x_{i}+x_{j}-1 
\end{split}
\end{equation}\\
\\
Eventually, the MIP formulation of the original QUBO problem reads:

\begin{equation}
    \sum_{i}\sum_{j \neq i}Q_{ij}y_{ij}+\sum_{i}Q_{ii}x_{i} \ ,
\end{equation}
subject to the above constraints (Eq. (\ref{eq:mip_constraint})).\\

Having obtained the MIP form of the QUBO problem, Gurobi can efficiently solve the problem using the branch-and-bound algorithm \cite{morrison2016branch,boyd2007branch,narendra1977branch}. The branch-and-bound algorithm begins by relaxing the binary variables to continuous ones and then solving the resulting LP with a mixture of simplex and barrier methods \cite{klee1972good,gill1986projected}. The optimal value of the LP serves as the lower bound for the root node, while the upper bound is the objective value of a feasible solution obtained via heuristics (such as rounding the relaxed solutions). Next, two branches of a variable $x$ from the optimal solution are extracted to form two subproblems with conditions $x=0$ and $x=1$. These subproblems are solved again for the lower and upper bounds. If a lower upper bound is found, the current best upper bound is updated. Conversely, if the lower bound of a certain node exceeds the current best upper bound, the node is pruned and no further exploration is necessary along the associated branch. We can continue with the branching and bounding process by repeating the above steps until all branches are pruned or explored. Eventually, the optimal solution corresponds to the feasible solution that gives rise to the lowest upper bound.

\subsection{D-Wave Advantage Quantum Annealer}
\label{section:DWave}

The D-Wave Advantage is currently the largest quantum annealer available, featuring over 5,000 qubits and connectivity of 15 couplers per qubit, facilitated by its Pegasus topology. The Hamiltonian of a quantum annealer can be described by the transverse-field Ising model:

\begin{equation}
    H_{Ising} =\sum_{i}h_{i}^{\prime}\sigma_{i}^{x}+\sum_{i}h_{i}\sigma_{i}^{z}+\sum_{<i,j>}J_{ij}\sigma_{i}^{z}\sigma_{j}^{z} \ ,
\end{equation}
where the eigenvalues of $\sigma_{i}^{z}$ are $\pm 1$. Since the mathematical formulation of the Ising model resembles that of a QUBO cost function, the two can be mapped with a linear translation between the binary variables and the Pauli-Z operator, where $x_{i} = (1+\sigma^{z}_{i})/2$. Due to this similarity, quantum annealing is regarded as a promising candidate for QUBO optimization. According to the adiabatic theorem, if we change the Hamiltonian of a quantum system adiabatically, we can bring the system from the ground state (or any eigenstate) of the initial Hamiltonian $H_{I}$ to the ground state (instantaneous eigenstate) of the target Hamiltonian $H_{f}$. The evolution protocol of an adiabatic process can be expressed as a time-changing linear combination of $H_{I}$ and $H_{f}$:

\begin{equation}\label{eq:adiabatic}
    H(t) = (1-\frac{t}{T})H_{I}+\frac{t}{T}H_{f} \ \ , \ \ t \in [0,T]
\end{equation}\\

 Inspired by the adiabatic theorem, quantum annealing aims to retrieve the optimal solution of a QUBO problem by evolving quantum states and leveraging the properties of quantum tunneling and superposition. To achieve this, the QUBO cost function is first encoded as the target Hamiltonian $H_{f}$ by mapping the linear terms to the longitudinal fields and the quadratic terms to the interaction of the Ising Hamiltonian. In this manner, preparing the ground state of the target Hamiltonian is equivalent to maneuvering through the energy landscape toward the minimal solution. In principle, the annealing protocol of Advantage follows Eq.(\ref{eq:adiabatic}), where $H_{I}=\sum_{i}h_{i}^{\prime}\sigma_{i}^{x}$ represents the transverse field that gradually attenuates with time. The initial state is the uniform superposition $\otimes_{i} (\ket{+z}-\ket{-z})$, which corresponds to the ground state of $H_{I}$.

 \subsection{ Fujitsu Digital Annealer}
 \label{section:Fujitsu}

Digital annealing is a quantum-inspired algorithm implemented on specialized classical hardware. The backbone of digital annealing is simulated annealing except that the algorithm is enhanced by optimally-designed hardware and has more sophisticated add-on features. Unlike quantum annealing, digital annealers are less susceptible to noise and offer greater scalability, as they rely on digital circuit technology rather than quantum computing platforms. In simulated annealing, the system is initialized in a random configuration of binary variables. The annealing process follows a temperature schedule that gradually reduces from a high value to almost zero. At each temperature $T$, several iterations of solution exploration are carried out. In each iteration, a binary variable is inverted, and a criterion is used to evaluate whether this new solution should be accepted. If the new solution has a lower energy than the previous one, this solution would be taken. Otherwise, the current solution would only be replaced with a Boltzmann probability $e^{-\Delta E/T}$, where $\Delta E > 0$ is the energy difference between the new candidate solution and the current one. As the temperature decreases, the probability of climbing over energy barriers to other parts of the solution space also decreases, and the system gradually converges to a stable state. The Fujitsu digital annealer is equipped with two unique features on top of the conventional simulated annealing. First, the solver is capable of detecting local minima and applying an offset energy when calculating the Boltzmann probability to encourage escape from the minima. Such a feature mimics the tunneling effect of quantum annealing, increasing the likelihood of finding the optimal solution. Second, the digital annealer has full connectivity between all bits in the device, enabling parallel processing of bit inversion as opposed to the sequential trial of simulated annealing. The parallelism of digital annealer emulates quantum superposition and accelerates the convergence to optimal solutions when dealing with large QUBO problems.

\section{Post-processing of FNN}
 \label{section:postprocessing}

After obtaining a presumably good solution vector from the FNN, a post-processing technique similar to classical simulated annealing can be applied to further enhance the performance of the FNN optimizer without introducing significant computational overhead. This method features a temperature parameter that regulates the trade-off between precision of modifications and processing time. The post-processing is composed of ten rounds with decreasing temperatures $T(r)$ proportional to the cost value of the output vector $C_{FNN}$, and inversely proportional to the processing round $r \in \{1, 2, \ldots, 10\}$ and the problem size $n$. The temperature profile is given by:

\begin{equation}
    T(r) = a\frac{|C_{FNN}|}{nr} \ ,
\end{equation}
where $a$ is some tunable constant, which we chose to be $1$. During each round, the algorithm iterates over every bit of the solution vector, inverting one bit at a time to probe for better solutions. If a lower cost value is found, the current best solution is updated. Conversely, if a higher cost value is produced, the candidate solution is retained with Boltzmann probability $\exp(-\Delta E/T(r))$, where $\Delta E = C_{candidate}-C_{best}$ is the difference between the cost values of the candidate and the current best solution. The process of iterating and inverting bits continues for the retained candidate solution, but solutions yielding higher cost values are no longer accepted—only better solutions are considered. By repeating the process for ten rounds with decreasing acceptance probabilities, we increase the likelihood of obtaining or approaching the optimal solution.\\

Empirically, this post-processing technique did not noticeably impact the computation time for QUBO problems with fewer than 80 variables. For 200-variable problems, the increase in runtime was also within 10\%. While iterating over all elements of the solution vector entails at least a linear time scaling, the acceptance probability $\exp(-\Delta E/T(r))$ dictates how much the computation might be extended, which is upper-bounded by quadratic scaling. In theory, a very high acceptance probability approximates a brute-force search over every pair of two-bit inversions for each round, leading to the most accurate correction in this setup at the cost of the longest post-processing time. Therefore, a proper choice of the temperature profile is key to balancing both efficiency and accuracy, as the temperature directly influences the value of the acceptance probability.


\bibliographystyle{iopart-num.bst}
\bibliography{IP}

\providecommand{\noopsort}[1]{}\providecommand{\singleletter}[1]{#1}%
\providecommand{\newblock}{}
\begin{thebibliography}{100}
\expandafter\ifx\csname url\endcsname\relax
  \def\url#1{{\tt #1}}\fi
\expandafter\ifx\csname urlprefix\endcsname\relax\def\urlprefix{URL }\fi
\providecommand{\eprint}[2][]{\url{#2}}

\bibitem{bangert2012optimization}
Bangert P 2012 {\em Optimization for Industrial Problems\/} (Springer Science \& Business Media)

\bibitem{garey1979computers}
Garey M~R and Johnson D~S 1979 {\em Computers and Intractability\/} vol 174 (Freeman San Francisco)

\bibitem{lucas2014ising}
Lucas A 2014 {\em Frontiers in Physics\/} {\bf 2} 5

\bibitem{gurobi}
{Gurobi Optimization, LLC} 2023 {Gurobi Optimizer Reference Manual} \urlprefix\url{https://www.gurobi.com}

\bibitem{glover2019quantum}
Glover F, Kochenberger G and Du Y 2019 {\em 4OR-Q Journal of Operations Research\/} {\bf 17} 335--371

\bibitem{lewis2017quadratic}
Lewis M and Glover F 2017 {\em Networks\/} {\bf 70} 79--97

\bibitem{commander2009maximum}
Commander C~W 2009 Maximum cut problem, {MAX}-cut {\em Encyclopedia of Optimization\/} vol~2 (Springer)

\bibitem{basagni2001finding}
Basagni S 2001 {\em Telecommunication Systems\/} {\bf 18} 155--168

\bibitem{matai2010traveling}
Matai R, Singh S and Mittal M~L 2010 Traveling salesman problem: An overview of applications, formulations, and solution approaches {\em Traveling Salesman Problem\/} ed Donald D (Rijeka: IntechOpen) chap~1

\bibitem{black1992global}
Black F and Litterman R 1992 {\em Financial Analysts Journal\/} {\bf 48} 28--43

\bibitem{yu2008workflow}
Yu J, Buyya R and Ramamohanarao K 2008 Workflow scheduling algorithms for grid computing {\em Metaheuristics for Scheduling in Distributed Computing Environments\/} (Springer) pp 173--214

\bibitem{nambu2022rejection}
Nambu Y 2022 {\em IEEE Access\/} {\bf 10} 84279--84301

\bibitem{nakano2023diverse}
Nakano K, Takafuji D, Ito Y, Yazane T, Yano J, Ozaki S, Katsuki R and Mori R 2023 Diverse adaptive bulk search: a framework for solving {QUBO} problems on multiple {GPUs} {\em 2023 IEEE International Parallel and Distributed Processing Symposium Workshops (IPDPSW)\/} (IEEE) pp 314--325

\bibitem{glover1993user}
Glover F, Taillard E and Taillard E 1993 {\em Annals of Operations Research\/} {\bf 41} 1--28

\bibitem{kirkpatrick1983optimization}
Kirkpatrick S, Gelatt~Jr C~D and Vecchi M~P 1983 {\em Science\/} {\bf 220} 671--680

\bibitem{guerreschi2021solving}
Guerreschi G~G 2021 {\em arXiv:2101.07813\/}

\bibitem{morrison2016branch}
Morrison D~R, Jacobson S~H, Sauppe J~J and Sewell E~C 2016 {\em Discrete Optimization\/} {\bf 19} 79--102

\bibitem{nabli2009overview}
Nabli H 2009 {\em Applied Mathematics and Computation\/} {\bf 210} 479--489

\bibitem{schuetz2022combinatorial}
Schuetz M~J, Brubaker J~K and Katzgraber H~G 2022 {\em Nature Machine Intelligence\/} {\bf 4} 367--377

\bibitem{gabor2020insights}
Gabor T, Feld S, Safi H, Phan T and Linnhoff-Popien C 2020 Insights on training neural networks for {QUBO} tasks {\em Proceedings of the IEEE/ACM 42nd International Conference on Software Engineering Workshops\/} (IEEE) pp 436--441

\bibitem{merz1999genetic}
Merz P and Freisleben B 1999 Genetic algorithms for binary quadratic programming {\em Proceedings of the Genetic and Evolutionary Computation Conference\/} vol~1 (Morgan Kaufmann Orlando, FL) pp 417--424

\bibitem{boros2007local}
Boros E, Hammer P~L and Tavares G 2007 {\em Journal of Heuristics\/} {\bf 13} 99--132

\bibitem{ohzeki2011quantum}
Ohzeki M and Nishimori H 2011 {\em Journal of Computational and Theoretical Nanoscience\/} {\bf 8} 963--971

\bibitem{kadowaki1998quantum}
Kadowaki T and Nishimori H 1998 {\em Physical Review E\/} {\bf 58} 5355

\bibitem{farhi2000quantum}
Farhi E, Goldstone J, Gutmann S and Sipser M 2000 {\em arXiv:quant-ph/0001106\/}

\bibitem{farhi2014quantum}
Farhi E, Goldstone J and Gutmann S 2014 {\em arXiv:1411.4028\/}

\bibitem{grover1996fast}
Grover L~K 1996 A fast quantum mechanical algorithm for database search {\em Proceedings of the Twenty-Eighth Annual ACM Symposium on Theory of Computing\/} (ACM) pp 212--219

\bibitem{venegas2012quantum}
Venegas-Andraca S~E 2012 {\em Quantum Information Processing\/} {\bf 11} 1015--1106

\bibitem{ebadi2022quantum}
Ebadi S, Keesling A, Cain M, Wang T~T, Levine H, Bluvstein D, Semeghini G, Omran A, Liu J, Samajdar R {\em et~al.\/} 2022 {\em Science\/} {\bf 376} 1209--1215

\bibitem{motta2020determining}
Motta M, Sun C, Tan A~T, O’Rourke M~J, Ye E, Minnich A~J, Brandao F~G and Chan G~K~L 2020 {\em Nature Physics\/} {\bf 16} 205--210

\bibitem{del2013shortcuts}
Del~Campo A 2013 {\em Physical Review Letters\/} {\bf 111} 100502

\bibitem{moll2018quantum}
Moll N, Barkoutsos P, Bishop L~S, Chow J~M, Cross A, Egger D~J, Filipp S, Fuhrer A, Gambetta J~M, Ganzhorn M {\em et~al.\/} 2018 {\em Quantum Science and Technology\/} {\bf 3} 030503

\bibitem{egger2021warm}
Egger D~J, Mare{\v{c}}ek J and Woerner S 2021 {\em Quantum\/} {\bf 5} 479

\bibitem{orus2019quantum}
Or{\'u}s R, Mugel S and Lizaso E 2019 {\em Reviews in Physics\/} {\bf 4} 100028

\bibitem{yarkoni2022quantum}
Yarkoni S, Raponi E, B{\"a}ck T and Schmitt S 2022 {\em Reports on Progress in Physics\/} {\bf 85} 104001

\bibitem{abbas2023quantum}
Abbas A, Ambainis A, Augustino B, B{\"a}rtschi A, Buhrman H, Coffrin C, Cortiana G, Dunjko V, Egger D~J, Elmegreen B~G {\em et~al.\/} 2023 {\em arXiv:2312.02279\/}

\bibitem{boyd2007branch}
Boyd S and Mattingley J 2007 {\em Notes for EE364b, Stanford University\/} {\bf 2006} 07

\bibitem{narendra1977branch}
Narendra and Fukunaga 1977 {\em IEEE Transactions on Computers\/} {\bf 100} 917--922

\bibitem{cplex2009v12}
Cplex I~I 2009 {\em International Business Machines Corporation\/} {\bf 46} 157

\bibitem{van1987simulated}
Van~Laarhoven P~J, Aarts E~H, van Laarhoven P~J and Aarts E~H 1987 {\em Simulated Annealing\/} (Springer)

\bibitem{nelder1965simplex}
Nelder J~A and Mead R 1965 {\em The Computer Journal\/} {\bf 7} 308--313

\bibitem{alaa2024efficient}
Alaa El-Din K, Alexander O, Frasinski L {\em et~al.\/} 2024 {\em Scientific Reports\/} {\bf 14} 7267

\bibitem{de2004machine}
De~La~Calleja J and Fuentes O 2004 {\em Monthly Notices of the Royal Astronomical Society\/} {\bf 349} 87--93

\bibitem{carleo2019machine}
Carleo G, Cirac I, Cranmer K, Daudet L, Schuld M, Tishby N, Vogt-Maranto L and Zdeborov{\'a} L 2019 {\em Reviews of Modern Physics\/} {\bf 91} 045002

\bibitem{pichler2018computational}
Pichler H, Wang S~T, Zhou L, Choi S and Lukin M~D 2018 {\em arXiv:1809.04954\/}

\bibitem{goswami2023solving}
Goswami K, Mukherjee R, Ott H and Schmelcher P 2023 {\em arXiv:2308.07798\/}

\bibitem{peruzzo2014variational}
Peruzzo A, McClean J, Shadbolt P, Yung M~H, Zhou X~Q, Love P~J, Aspuru-Guzik A and O’Brien J~L 2014 {\em Nature Communications\/} {\bf 5} 4213

\bibitem{bravo2022quantum}
Bravo R~A, Najafi K, Gao X and Yelin S~F 2022 {\em PRX Quantum\/} {\bf 3} 030325

\bibitem{bosch2023neural}
Bosch S, Kiani B, Yang R, Lupascu A and Lloyd S 2023 {\em arXiv:2308.06807\/}

\bibitem{yang2023analog}
Yang R, Bosch S, Kiani B, Lloyd S and Lupascu A 2023 {\em Physical Review Applied\/} {\bf 19} 054023

\bibitem{he2024quantum}
He H 2024 {\em arXiv:2402.14036\/}

\bibitem{adachi2015application}
Adachi S~H and Henderson M~P 2015 {\em arXiv:1510.06356\/}

\bibitem{higham2023quantum}
Higham C~F and Bedford A 2023 {\em Scientific Reports\/} {\bf 13} 3939

\bibitem{goswami2024quantum}
Goswami K, Schmelcher P and Mukherjee R 2024 {\em Bulletin of the American Physical Society\/} {\bf 69} 1--5

\bibitem{goswami2024solving}
Goswami K, Veereshi G~A, Schmelcher P and Mukherjee R 2024 {\em arXiv:2407.17207\/}

\bibitem{bebis1994feed}
Bebis G and Georgiopoulos M 1994 {\em IEEE Potentials\/} {\bf 13} 27--31

\bibitem{eldan2016power}
Eldan R and Shamir O 2016 The power of depth for feedforward neural networks {\em Conference on Learning Theory\/} (PMLR) pp 907--940

\bibitem{fine2006feedforward}
Fine T~L 2006 {\em Feedforward Neural Network Methodology\/} (Springer Science \& Business Media)

\bibitem{hecht1992theory}
Hecht-Nielsen R 1992 Theory of the backpropagation neural network {\em Neural Networks for Perception\/} (Elsevier) pp 65--93

\bibitem{werbos1990backpropagation}
Werbos P~J 1990 {\em Proceedings of the IEEE\/} {\bf 78} 1550--1560

\bibitem{amari1993backpropagation}
Amari S~i 1993 {\em Neurocomputing\/} {\bf 5} 185--196

\bibitem{lecun1988theoretical}
LeCun Y, Touretzky D~S, Hinton G~E and Sejnowski T~J 1988 A theoretical framework for back-propagation {\em Proceedings of the 1988 Connectionist Models Summer School\/} vol~1 (Morgan Kaufmann) pp 21--28

\bibitem{zhang2018improved}
Zhang Z 2018 Improved {Adam} optimizer for deep neural networks {\em 2018 IEEE/ACM 26th International Symposium on Quality of Service (IWQoS)\/} IEEE (IEEE) pp 1--2

\bibitem{zou2019sufficient}
Zou F, Shen L, Jie Z, Zhang W and Liu W 2019 A sufficient condition for convergences of {Adam} and {RMSprop} {\em Proceedings of the IEEE/CVF Conference on Computer Vision and Pattern Recognition\/} (IEEE) pp 11127--11135

\bibitem{zinkevich2010parallelized}
Zinkevich M, Weimer M, Li L and Smola A 2010 {\em Advances in Neural Information Processing Systems\/} {\bf 23} 2597--2605

\bibitem{ketkar2017stochastic}
Ketkar N and Ketkar N 2017 {\em Deep Learning with Python: A Hands-On Introduction\/}  113--132

\bibitem{kim2022rydberg}
Kim M, Kim K, Hwang J, Moon E~G and Ahn J 2022 {\em Nature Physics\/} {\bf 18} 755--759

\bibitem{lanthaler2023rydberg}
Lanthaler M, Dlaska C, Ender K and Lechner W 2023 {\em Physical Review Letters\/} {\bf 130} 220601

\bibitem{qiu2020programmable}
Qiu X, Zoller P and Li X 2020 {\em PRX Quantum\/} {\bf 1} 020311

\bibitem{nguyen2023quantum}
Nguyen M~T, Liu J~G, Wurtz J, Lukin M~D, Wang S~T and Pichler H 2023 {\em PRX Quantum\/} {\bf 4} 010316

\bibitem{goswami2024loqal}
Goswami K, Mukherjee R, Ott H and Schmelcher P 2024 {\em Physical Review Research\/} {\bf 6} 023031

\bibitem{kaufman2021quantum}
Kaufman A~M and Ni K~K 2021 {\em Nature Physics\/} {\bf 17} 1324--1333

\bibitem{zeiher2016many}
Zeiher J, Van~Bijnen R, Schau{\ss} P, Hild S, Choi J~y, Pohl T, Bloch I and Gross C 2016 {\em Nature Physics\/} {\bf 12} 1095--1099

\bibitem{browaeys2020many}
Browaeys A and Lahaye T 2020 {\em Nature Physics\/} {\bf 16} 132--142

\bibitem{bluvstein2021controlling}
Bluvstein D, Omran A, Levine H, Keesling A, Semeghini G, Ebadi S, Wang T~T, Michailidis A~A, Maskara N, Ho W~W {\em et~al.\/} 2021 {\em Science\/} {\bf 371} 1355--1359

\bibitem{ates2007many}
Ates C, Pohl T, Pattard T and Rost J~M 2007 {\em Physical Review A\/} {\bf 76} 013413

\bibitem{honer2010collective}
Honer J, Weimer H, Pfau T and B{\"u}chler H~P 2010 {\em Physical Review Letters\/} {\bf 105} 160404

\bibitem{gaetan2009observation}
Ga{\"e}tan A, Miroshnychenko Y, Wilk T, Chotia A, Viteau M, Comparat D, Pillet P, Browaeys A and Grangier P 2009 {\em Nature Physics\/} {\bf 5} 115--118

\bibitem{johnson2010interactions}
Johnson J and Rolston S 2010 {\em Physical Review A\/} {\bf 82} 033412

\bibitem{weber2017calculation}
Weber S, Tresp C, Menke H, Urvoy A, Firstenberg O, B{\"u}chler H~P and Hofferberth S 2017 {\em Journal of Physics B: Atomic, Molecular and Optical Physics\/} {\bf 50} 133001

\bibitem{gallagher2008dipole}
Gallagher T~F and Pillet P 2008 {\em Advances in Atomic, Molecular, and Optical Physics\/} {\bf 56} 161--218

\bibitem{urban2009observation}
Urban E, Johnson T~A, Henage T, Isenhower L, Yavuz D, Walker T and Saffman M 2009 {\em Nature Physics\/} {\bf 5} 110--114

\bibitem{schuld2019evaluating}
Schuld M, Bergholm V, Gogolin C, Izaac J and Killoran N 2019 {\em Physical Review A\/} {\bf 99} 032331

\bibitem{crooks2019gradients}
Crooks G~E 2019 {\em arXiv:1905.13311\/}

\bibitem{wierichs2022general}
Wierichs D, Izaac J, Wang C and Lin C~Y~Y 2022 {\em Quantum\/} {\bf 6} 677

\bibitem{banchi2021measuring}
Banchi L and Crooks G~E 2021 {\em Quantum\/} {\bf 5} 386

\bibitem{mcgeoch2020d}
McGeoch C and Farr{\'e} P 2020 {\em D-Wave Systems Inc., Burnaby, BC, Canada, Tech. Rep\/}

\bibitem{matsubara2020digital}
Matsubara S, Takatsu M, Miyazawa T, Shibasaki T, Watanabe Y, Takemoto K and Tamura H 2020 Digital annealer for high-speed solving of combinatorial optimization problems and its applications {\em 2020 25th Asia and South Pacific Design Automation Conference (ASP-DAC)\/} IEEE (IEEE) pp 667--672

\bibitem{ullah2022quantum}
Ullah M~H, Eskandarpour R, Zheng H and Khodaei A 2022 {\em IET Generation, Transmission \& Distribution\/} {\bf 16} 4239--4257

\bibitem{neukart2017traffic}
Neukart F, Compostella G, Seidel C, Von~Dollen D, Yarkoni S and Parney B 2017 {\em Frontiers in ICT\/} {\bf 4} 29

\bibitem{tatsumura2023real}
Tatsumura K, Hidaka R, Nakayama J, Kashimata T and Yamasaki M 2023 {\em IEEE Access\/} {\bf 11} 120023--120033

\bibitem{ramirez2020integrating}
Ram{\'\i}rez J~G~C 2020 {\em Quarterly Journal of Emerging Technologies and Innovations\/} {\bf 5} 11--22

\bibitem{wang2022integrating}
Wang H, Wang W, Liu Y and Alidaee B 2022 {\em IEEE Access\/} {\bf 10} 75908--75917

\bibitem{wu2019hyperparameter}
Wu J, Chen X~Y, Zhang H, Xiong L~D, Lei H and Deng S~H 2019 {\em Journal of Electronic Science and Technology\/} {\bf 17} 26--40

\bibitem{victoria2021automatic}
Victoria A~H and Maragatham G 2021 {\em Evolving Systems\/} {\bf 12} 217--223

\bibitem{snoek2015scalable}
Snoek J, Rippel O, Swersky K, Kiros R, Satish N, Sundaram N, Patwary M, Prabhat M and Adams R 2015 Scalable {Bayesian} optimization using deep neural networks {\em International Conference on Machine Learning\/} PMLR (PMLR) pp 2171--2180

\bibitem{jospin2022hands}
Jospin L~V, Laga H, Boussaid F, Buntine W and Bennamoun M 2022 {\em IEEE Computational Intelligence Magazine\/} {\bf 17} 29--48

\bibitem{mukherjee2020preparation}
Mukherjee R, Sauvage F, Xie H, L{\"o}w R and Mintert F 2020 {\em New Journal of Physics\/} {\bf 22} 075001

\bibitem{hatano2005finding}
Hatano N and Suzuki M 2005 Finding exponential product formulas of higher orders {\em Quantum Annealing and Other Optimization Methods\/} (Springer) pp 37--68

\bibitem{suzuki1976generalized}
Suzuki M 1976 {\em Communications in Mathematical Physics\/} {\bf 51} 183--190

\bibitem{delorme2004grasp}
Delorme X, Gandibleux X and Rodriguez J 2004 {\em European Journal of Operational Research\/} {\bf 153} 564--580

\bibitem{balas1976set}
Balas E and Padberg M~W 1976 {\em SIAM Review\/} {\bf 18} 710--760

\bibitem{lawler1963quadratic}
Lawler E~L 1963 {\em Management Science\/} {\bf 9} 586--599

\bibitem{korf2009multi}
Korf R~E 2009 Multi-way number partitioning {\em Twenty-First International Joint Conference on Artificial Intelligence\/} (AAAI Press) pp 1010--1015

\bibitem{gerlach2024investigating}
Gerlach T and M{\"u}cke S 2024 Investigating the relation between problem hardness and {QUBO} properties {\em International Symposium on Intelligent Data Analysis\/} (Springer) pp 171--182

\bibitem{mehta2022hardness}
Mehta V, Jin F, Michielsen K and De~Raedt H 2022 {\em Frontiers in Physics\/} {\bf 10} 956882

\bibitem{dantzig2002linear}
Dantzig G~B 2002 {\em Operations Research\/} {\bf 50} 42--47

\bibitem{ignizio1994linear}
Ignizio J~P and Cavalier T~M 1994 {\em Linear Programming\/} (Prentice-Hall, Inc.)

\bibitem{wolsey2007mixed}
Wolsey L~A 2007 {\em Wiley Encyclopedia of Computer Science and Engineering\/}  1--10

\bibitem{smith2008tutorial}
Smith J~C and Taskin Z~C 2008 {\em Optimization in Medicine and Biology\/}  521--548

\bibitem{klee1972good}
Klee V and Minty G~J 1972 {\em Inequalities\/} {\bf 3} 159--175

\bibitem{gill1986projected}
Gill P~E, Murray W, Saunders M~A, Tomlin J~A and Wright M~H 1986 {\em Mathematical Programming\/} {\bf 36} 183--209

\end{thebibliography}

\end{document}